\documentclass[usenatbib]{mn2e}
\usepackage{amsmath}
\usepackage{graphicx,times}
\usepackage{bm,url}
\usepackage{natbib}
\newcommand{\EQ}{\begin{equation}}
\newcommand{\EN}{\end{equation}}

\newcommand{\mkG}{\,{\rm \mu G}}

\newcommand{\befone}{
  \begin{figure*}
  \centering
  \begin{minipage}{\textwidth}
  }
\newcommand{\eefone}{\end{minipage}\end{figure*}}

\def\figthree@scaling{0.35}
\def\showthree#1#2#3{
  \centering
  \leavevmode
  \includegraphics[height=\figthree@scaling\columnwidth]{#1}
  \includegraphics[height=\figthree@scaling\columnwidth]{#2}
  \includegraphics[height=\figthree@scaling\columnwidth]{#3}
}

\def\figtwo@scaling{0.44}
\def\showfour#1#2#3#4{
  \centering
  \leavevmode
  \includegraphics[width=\figtwo@scaling\linewidth]{#1} \hfil
  \includegraphics[width=\figtwo@scaling\linewidth]{#2} \hfil
  \includegraphics[width=\figtwo@scaling\linewidth]{#3} \hfil
  \includegraphics[width=\figtwo@scaling\linewidth]{#4}
}

\def\figtwo@scaling{0.44}
\def\showsix#1#2#3#4#5#6{
  \centering
  \leavevmode
  \includegraphics[width=\figtwo@scaling\linewidth]{#1} \hfil
  \includegraphics[width=\figtwo@scaling\linewidth]{#2} \hfil
  \includegraphics[width=\figtwo@scaling\linewidth]{#3} \hfil
  \includegraphics[width=\figtwo@scaling\linewidth]{#4} \hfil
  \includegraphics[width=\figtwo@scaling\linewidth]{#5} \hfil
   \includegraphics[width=\figtwo@scaling\linewidth]{#6} 
}

\title[Magnetic field amplification during gravitational collapse]%
{Magnetic field amplification during gravitational collapse - Influence of initial conditions 
on dynamo evolution and saturation}
\author[S.~Sur, C.~Federrath, D.~R.~G Schleicher, R.~Banerjee \& R.~Klessen]%
{Sharanya Sur$^{1,2,3}$, Christoph Federrath$^{1,4}$, Dominik R.~G.~Schleicher$^{5}$, \and
Robi Banerjee$^{6}$ and Ralf S.~Klessen$^{1}$
\thanks{E-mail: ssur06@gmail.com(SS)}\\
$^{1}$Zentrum f\"ur Astronomie der Universit\"at Heidelberg, Institut f\"ur 
Theoretische Astrophysik, Albert-Ueberle-Str.~2, 69120 Heidelberg, Germany\\
$^{2}$Inter-University Centre for Astronomy \& Astrophysics, Post Bag 4, Ganeshkhind, Pune - 411007, India\\
$^{3}$Raman Research Institute, C.~V.~ Raman Avenue, Sadashivnagar, Bangalore, 560080, India\\
$^{4}$Monash Center for Astrophysics (MoCA), School of Mathematical Sciences, Monash University, Vic 3800, Australia\\
$^{5}$Georg-August-Universit\"at, Institut f\"ur Astrophysik, Friedrich-Hund-Platz 1, G\"ottingen, Germany\\
$^{6}$Hamburger Sternwarte, Gojenbergsweg 112, 21029 Hamburg, Germany \\
}

\date{}
\begin{document}

\pagerange{\pageref{firstpage}--\pageref{lastpage}} \pubyear{2011}

\maketitle
 
%---------------------------------------------------------------------
\begin{abstract}
We study the influence of initial conditions on the magnetic field amplification during the collapse 
of a magnetised gas cloud. We focus on the dependence of the growth and saturation 
level of the dynamo generated field on the turbulent properties of the collapsing cloud. 
In particular, we explore the effect of varying the initial strength and injection scale of turbulence and the 
initial uniform rotation of the collapsing magnetised cloud. In order to follow the evolution of the magnetic 
field in both the kinematic and the nonlinear regime, we choose an initial field strength of $\simeq 1\,\mkG$ with 
the magnetic to kinetic energy ratio, $E_{\rm m}/E_{\rm k} \sim 10^{-4}$. Both gravitational compression 
and the small-scale dynamo initially amplify the magnetic field. Further into the evolution, the dynamo-generated 
magnetic field saturates but the total magnetic field continues to grow because of compression. The saturation 
of the small-scale dynamo is marked by a change in the slope of $B/\rho^{2/3}$ and by a 
shift in the peak of the magnetic energy spectrum from small scales to larger scales. For the range of initial 
Mach numbers explored in this study, the dynamo growth rate increases as the Mach number increases from 
$v_{\rm rms}/c_{\rm s}\sim 0.2$ to $0.4$ and then starts decreasing from $v_{\rm rms}/c_{\rm s}\sim 1.0$. We 
obtain saturation values of $E_{\rm m}/E_{\rm k} = 0.2 - 0.3$ for these runs. Simulations with different initial 
injection scales of turbulence also show saturation at similar levels. For runs with different initial rotation of the 
cloud, the magnetic energy saturates at  $E_{\rm m}/E_{\rm k}\sim 0.2 - 0.4$ of the equipartition value. The 
overall saturation level of the magnetic energy, obtained by varying the initial conditions are in agreement with 
previous analytical and numerical studies of small-scale dynamo action where turbulence is driven by an 
external forcing instead of gravitational collapse. 
\end{abstract}
\label{firstpage}

%----------------------------------------------------------------------------------
\section{Introduction}
\label{Intro}
Magnetic fields are ubiquitous in astrophysical systems and their study forms an active 
area of research today. Radio observations over the last few decades have revealed that 
galaxies and galaxy clusters host magnetic fields. The total magnetic field strength in nearby 
spiral galaxies is $\sim 9 - 15\,\mkG$ \citep{BH96, Beck04} while observations of cluster
magnetic fields show that fields are at the $\mkG$ level, with values up to tens of $\mkG$ 
at the center of cooling core clusters \citep{CT02, GF04}. Recent 
observations also point to the existence of magnetic fields in the high-redshift universe 
\citep{B+08}. One plausible mechanism of the origin of such magnetic fields is the {\it dynamo} 
process where energy in the turbulent fluid motions is tapped to amplify the magnetic field. 
Turbulence is ubiquitous in all astrophysical systems ranging from protostellar accretion disks 
in individual star forming clouds to the interstellar medium (ISM) in galaxies and possibly also 
in the gaseous media of galaxy clusters and cosmic filaments. In fluid dynamics, turbulence is 
described as a flow regime characterised by chaotic motions and involves the cascade of energy
from the scale of the largest eddy to the smallest scale eddy. 
In particular, the {\it small-scale} dynamo process \citep{K68, Vainstein82, SBK02, BC04, Sur+10, 
Federrath+11, SSFKB11} 
can lead to rapid amplification of initial seed magnetic fields \citep[see][for a review]{BS05}. 
Earlier work by \citet{BPSS94} proposed that such dynamo action is responsible for producing seed 
magnetic fields for the galactic large-scale dynamo. The same mechanism is found to amplify magnetic 
fields in galaxy clusters \citep{Dolag+99,SSH06,Xu+09, Xu+11} and also in the cosmological large-scale 
structure \citep{Ryu+08}. Mergers during cluster formation can also lead to turbulence and intense 
random vortical flows \citep{SM93,Kulsrud+97, BN98, Miniati+01,RS01,SSH06} capable of amplifying 
magnetic fields by the dynamo process \citep{SSH06}.

A potential application of small-scale dynamos concerns the formation of the first stars and the first 
galaxies where high-resolution simulations have revealed the ubiquity of turbulence in the early Universe 
suggesting that the primordial gas is highly turbulent \citep{ABN02, OsN07, WA07, YOH08, Greif+08,TAOs09}. 
This has strong implications for the exponential amplification of magnetic fields by the small-scale 
dynamo process during the formation of the first stars and in first galaxies 
\citep{Sch+10, Sur+10, Federrath+11, Turk+11}. High-resolution three-dimensional magnetohydrodynamics 
(MHD) simulations of collapsing magnetized primordial clouds by \citet{Sur+10} and \citet{Federrath+11} 
(hereafter Paper I and II respectively) show that the turbulence is driven by the gravitational collapse on 
scales of the order of the local Jeans length. This leads to an exponential growth of the magnetic field by 
random stretching, folding and twisting of the field lines. In the kinematic stage, the dynamo amplification 
occurs on the eddy-turnover time scale, $t_{\rm ed} = l/v$ where, $l$ and $v$ are the typical turbulent scale 
and turbulent velocity respectively. In a collapsing magnetized system, field amplification by the turbulent 
dynamo should occur on a time scale smaller than the free-fall time, $t_{\rm ff} \sim 1/\sqrt{G\,\rho}$ to 
enable field growth faster than the rate at which gravitational compression would amplify the field (Papers I and II). 
Here $\rho$ is the total mass density and $G$ is the gravitational constant. We note here that the efficiency 
of the dynamo process depends on the Reynolds number and is thus related to how well the turbulent 
motions are resolved in numerical simulations. Higher Reynolds number would yield faster field amplification. 
Indeed, in the limit of infinite magnetic Prandtl number (i.e., ${\rm Pr} = {\rm Rm}/{\rm Re} \rightarrow \infty$)
\footnote[1]{${\rm Rm}$ and ${\rm Re}$ are the magnetic and fluid Reynolds numbers respectively.}, 
\citet{SSFKB11} find a dependence of the growth rate on ${\rm Re}$, 
\EQ
\Gamma \sim f\,{\rm Re}^{(1-\xi)/(1+\xi)}
\EN
where $\xi = 1/3$ and $f=1.027$ or $\xi= 1/2$ and $f=0.184$ for Kolmogorov and Burgers turbulence respectively. 

However, the exorbitant computational costs associated with these simulations and the fact that current 
simulations largely underestimate the growth rate of the dynamo due to the modest Reynolds numbers 
achievable, restrict following the field evolution to only the kinematic phase of the dynamo. An important 
question concerning the magnetic field growth by the small-scale dynamo process is at what strength does 
the dynamo generated field saturate and what is the structure of those fields. Addressing this issue is important 
to obtain an estimate of the typical saturated field strengths to be expected in the protostellar cores. Saturation 
of the field is expected to occur when back reactions either via the Lorentz force \citep{S99, Scheko+04} or via 
non-ideal MHD effects such as ambipolar diffusion \citep{PG08} become important. In simulations of dynamo 
action with randomly forced turbulence in a box, saturation of the dynamo is achieved when the magnetic energy 
associated with the small-scale field grows to a fraction of the equipartition value \citep{HBD04, Scheko+04, BS05, 
Federrath+11b}. However, the exact meaning of saturation and saturated field strengths in a self-gravitating system 
is far from clear. In this case, the magnetic field is amplified by both turbulence and gravitational compression of 
the field lines. Therefore, saturation in a self-gravitating system involves also the gravitational energy and hence 
the ratio of magnetic to kinetic energies $E_{\rm m}/E_{\rm k}$, does not converge to a constant value i.e., 
$E_{\rm m}/E_{\rm k} \sim \rho^{1/3}$ (using the fact that the collapse speed $v(\rho)$ approaches a constant during 
the collapse \citep{Larson69}). Here, $E_{\rm m} \sim B^2\,V$ and $E_{\rm k} \sim \rho\,v^2\,V$ with $B$ being the 
magnetic field, $v, \rho$ and $V$ are the velocity, density and the volume respectively. Thus, we use $B\sim \rho^{2/3}$ 
to define dynamo saturation in our case. 

The dynamo amplification of magnetic fields is driven by turbulent fluid motions. Detailed parameter study
of small-scale dynamo action in driven turbulence in a box by \citet{Federrath+11b} show that the growth rate 
and the saturation level of the dynamo are sensitive to the Mach number and the turbulence injection mechanism. 
It is therefore crucial to explore how variations in the turbulent properties of a gas cloud influence the small-scale 
dynamo evolution and saturation in self-gravitating systems. In this spirit, we investigate the effects 
of environment on the gravitational collapse and magnetic field amplification in this paper. 
The key questions that we intend to address are the following: how does the collapse, growth and saturation 
level of the magnetic field depend on the initial strength and injection scale of the turbulent velocity? 
What information concerning dynamo saturation can be obtained from the spectra of magnetic fields?
We also seek to understand the collapse dynamics and the field amplification when the imposed 
uniform rotation of the cloud is varied. We address these questions using the initial conditions described in 
Papers I and II focussing on the collapse of a primordial gas cloud. 

An important issue when studying such systems is the choice of the initial field strength. The field strengths 
obtained from either cosmological processes like Inflation or phase transition mechanisms or from astrophysical 
mechanisms are weak and have large uncertainties \citep[see][for a review]{GR01}. Recent FERMI observations 
of TeV-blazars by \citet{NV10} and \citet{Tavecchio+10} have however reported a lower bound of about 
$(10^{-15} - 10^{-16} )\,{\rm Gauss}$ for the primordial field. Numerical simulations starting with such 
weak values for the initial field strength render it almost impossible to probe the saturation level of the dynamo 
generated fields due to the $\sim {\rm Re}^{1/2}$ scaling of the magnetic field growth rate (therefore requiring 
higher numerical resolution) and the high computational costs associated. Therefore, it is reasonable to start 
with a stronger seed field which allows us to probe the field amplification in both the kinematic and the nonlinear 
stage of the collapse. In all the simulations presented here, we therefore start with an initial field strength of 
$\sim 1\,\mkG$ with a ratio $E_{\rm m}/E_{\rm k} \sim 10^{-4}$ well below equipartition. 
In this sense, our simulations are to be viewed as controlled numerical experiments focussing on exploring the 
influence of certain parameters (e.g., the initial strength and injection scale of turbulence) on the collapse and 
magnetic field amplification. 

The paper is organized as follows. The numerical setup, initial conditions and the analysis methods of our 
simulations are outlined in Section 2. We present the results obtained from each of the three different parameter 
regimes i.e., varying the initial strength and injection scale of the turbulence, and the amount of initial rotation 
and in Section 3. Finally, we summarize and discuss the implications of our results in Section 4.

\section{Method}
To study the complex system involving self-gravity, turbulence and magnetic fields, we resort 
to high-resolution three-dimensional MHD simulations of a magnetised collapsing cloud using 
the adaptive-mesh refinement (AMR) technique.  
  
\subsection{Numerical setup and initial conditions}
The basic numerical setup is adopted from the one reported in Papers I and II. We focus on the gravitational collapse 
and magnetic field amplification of a dense gas cloud, using a simplified setup, where we assume a polytropic equation 
of state, $P\propto \rho^{\Gamma}$, which relates the pressure $P$ to the density $\rho$ with an exponent $\Gamma=1.1$ 
in the density range $\rho \sim (10^{-20} - 10^{-14})\,\rm{g}\,\,{\rm cm}^{-3}$. Note that this almost isothermal equation
of state is a good representation of the thermal behavior of the primordial gas at the densities 
considered here \citep{Omukai05, Clark+11}. The numerical simulations presented here were performed 
with the AMR code, FLASH2.5 \citep{Fryxell00}. We solve the equations of ideal MHD, including self-gravity 
with a refinement criterion guaranteeing that the Jeans length,
\EQ
\lambda_{\rm J} = \left(\frac{\pi c_{\rm s}^2}{G\,\rho}\right)^{1/2}
\EN
with sound speed $c_{\rm s}$, the gravitational constant $G$ and the density $\rho$ is always refined 
with a user defined number of cells. The applicability of ideal MHD in our simulations depends on 
the strength of the coupling between the gas and magnetic fields in primordial clouds. Work 
by \citet{MS04, MS07} focussing on detailed models of magnetic energy dissipation via Ohmic 
and ambipolar diffusion find that the ionization degree is sufficiently high in the primordial clouds to ensure 
a strong coupling between ions and neutrals, thereby maintaining perfect flux-freezing. This is specifically 
true for the primordial clouds which we address here. On the other hand, we expect a higher ionization degree 
and thus a more idealized situation, in the presence of additional radiation backgrounds in the later Universe.
However, non-ideal MHD effects may eventually become important at very high densities, as suggested by 
simulations of contemporary star formation \citep[e.g.,][]{HT08,DP09}. These effects are not included in the 
present calculations, but should be the subject of future studies. It is to be noted that since we use ideal MHD, 
the magnetic Prandtl number in our simulations is $\sim O(1)$. From earlier studies, we recall that the dynamo 
amplification requires a threshold resolution of about $30$ grid cells per Jeans length (see Papers I and II). 
Simulations performed with a resolution below 30 grid cells are unable to resolve the dynamo amplification 
of magnetic fields. We therefore perform simulations resolving the local Jeans length with a minimum of $64$ 
cells to a maximum of $128$ cells to explore the influence of initial conditions and the saturated field strengths 
on the small-scale dynamo generated field. 
We use the new HLL3R scheme for ideal MHD \citep{WFK11}, which employs a 3-wave approximate MHD 
Riemann solver \citep{BKW07,W09,BKW10}. The MHD scheme preserves physical states (e.g., positivity of 
mass density and pressure) by construction, and is highly efficient and accurate in modeling astrophysical 
MHD problems involving turbulence and shocks \citep{WFK11}. 

Similar to the studies reported in Papers I and II,  we model the gas cloud as an overdense Bonnor-Ebert (BE) 
sphere \citep{Eb55, Bonnor56} with a core density of $\rho_{\rm BE}=4.68\times10^{-20}\,\rm{g}\,\,{\rm cm}^{-3}$ 
at a temperature of $T=300\,{\rm K}$. Since the aim of this study is to investigate the influence of initial 
conditions on the evolution of the magnetic field, we choose our initial conditions in a way that allows us 
to perform a controlled numerical experiment. The initial random seed magnetic field of $1\,\mkG$ with a 
power-law dependence $P(k) \sim k^{-2}$ was constructed in Fourier space from the magnetic vector potential 
which automatically guarantees a divergence-free magnetic field. The turbulence is also modeled with an initial 
random velocity field with the same power-law dependence as for the initial magnetic field. The parameters of the 
suite of different simulations we perform is summarized in Table~\ref{sumsim}. The parameters $\alpha$ and $\beta$ 
quantify the ratio of initial turbulent energy and the rotational energy to the magnitude of the gravitational energy, 
respectively. Note that extremely high values of $\alpha$ may prevent the cloud from collapsing under its own 
gravity. Therefore, we choose initial turbulent velocities in the range $v_{\rm rms} \sim 0.2\,c_{\rm s} - 4\,c_{\rm s}$ 
such that $\alpha$ varies from a minimum of $9.3\times10^{-3}$ to a maximum of $\sim 3.723$. As for the 
initial injection scale of turbulence, we focus on two cases, one where the initial turbulence peaks on scales 
of the order of the initial Jeans length of the core ($l_{\rm inj}=0.7\,\lambda_{\rm J}$) and the other where it 
peaks on scales $l_{\rm inj}=0.17\,\lambda_{\rm J}$. The initial uniform rotation of the cloud is varied from 
$\beta = 0\%$ to $8\%$.

%%%%%%%%%%%%%%%%%%%%%%%%%%%%%%%%%%%%%%%%%%%%%%%%%%%%
\begin{table*}
\begin{minipage}{126mm}
%\begin{center}
\begin{tabular}{cccccc} \hline \hline 
Simulation & Resolution & Initial Turbulent &Initial Injection &  Initial turbulence & Initial rotation \\
Run & $[\# {\rm of \,cells}/\lambda_{\rm J}]$ & velocity $[v_{\rm rms}]$ & scale 
$[l_{\rm int}]$ & $\alpha=E_{\rm turb}/|E_{\rm grav}|$ & $\beta=E_{\rm rot}/|E_{\rm grav}|$
\\ \hline \hline
R64M0.2rot0 & 64 & 0.2\,$c_{\rm s}$ & 0.7\,$\lambda_{\rm J}$ & 9.30 $\times 10^{-3}$ & 0\% \\ \hline 
R64M0.4rot0 & 64 & 0.4\,$c_{\rm s}$ & 0.7\,$\lambda_{\rm J}$ & 3.72$\times 10^{-2}$ & 0\% \\ \hline
R64M1.0rot0 & 64 & 1.0\,$c_{\rm s}$ & 0.7\,$\lambda_{\rm J}$ & 0.232& 0\% \\ \hline
R64M2.0rot0 & 64 & 2.0\,$c_{\rm s}$ & 0.7\,$\lambda_{\rm J}$ & 0.930 & 0\% \\ \hline
R64M4.0rot0 & 64 & 4.0\,$c_{\rm s}$ & 0.7\,$\lambda_{\rm J}$ & 3.723& 0\% \\ \hline
R64M1.0rot0lj0.17 & 64 & $c_{\rm s}$ & 0.17\,$\lambda_{\rm J}$ & 0.233 & 0\% \\ \hline
R128M1.0rot0 & 128 & $c_{\rm s}$ & 0.7\,$\lambda_{\rm J}$ & 0.232& 0\%\\ \hline
R128M1.0rot4 & 128 & $c_{\rm s}$ & 0.7\,$\lambda_{\rm J}$ & 0.232& 4\% \\ \hline
R128M1.0rot8 & 128 & $c_{\rm s}$ & 0.7\,$\lambda_{\rm J}$ & 0.232& 8\% \\ \hline
R16M1.0rot4 & 16 & $c_{\rm s}$ & 0.7\,$\lambda_{\rm J}$ & 0.232& 4\% \\ \hline \hline
 \end{tabular}
%  \end{center}
\caption{Summary of the simulation runs. These fall in three broad categories : 
Runs with varying amplitude of the initial turbulence (runs R64M0.2rot0 to R64M4.0rot0)
followed by a run (R64M1.0rot0lj0.17) with a different injection scale of the initial turbulence.
Finally, runs with with different amounts of initial rotation (runs R16M1.0rot4 to R128M1.0rot8).}
\label{sumsim}
\end{minipage}
\end{table*}
%%%%%%%%%%%%%%%%%%%%%%%%%%%%%%%%%%%%%%%%%%%%%%%%%%%%
\subsection{Analysis in the collapsing frame of reference}
To understand the behavior of the system quantitatively, we need to follow its dynamical 
contraction in an appropriate frame of reference. First, we note that the physical time scale 
becomes progressively shorter during the collapse. We therefore define a dimensionless 
time coordinate $\tau$ (see Papers I and II),
\EQ
\label{eq:tau}
\tau = \int \frac{1}{t_\mathrm{ff}(t)}\,dt\,,
\EN
which is normalized in terms of the local free-fall time, 
\EQ
t_\mathrm{ff}(t)=\left(\frac{3\pi}{32\,G\rho_{\rm m}(t)}\right)^{1/2},
\EN
where $\rho_{\rm m}(t)$ is the mean density in the central Jeans volume, 
$V_\mathrm{J}=4\pi(\lambda_{\rm J}/2)^3/3$. If not otherwise stated, we obtain all dynamical 
quantities of interest within this contracting Jeans volume, which is centered on
the position of the maximum density. This approach enables us to study the turbulence 
and magnetic field amplification in the collapsing frame of reference.
We also note that for runs with different initial conditions, like varying degree of initial rotation, 
it is more meaningful to compare different simulations at the same central density rather than 
at the same $\tau$. This is because, runs with different initial conditions are in different phases 
of the collapse at any given $\tau$. Wherever possible, we therefore show plots of physical 
quantities as a function of the central mean density. With this in mind, we now proceed to discuss 
the effects of different initial conditions on the gravitational collapse and magnetic field amplification.  

\section{Results}

In this section, we present the results obtained from numerical simulations of magnetic fields in a 
collapsing environment. We report on the influence of varying the initial strength and injection scale 
of the turbulence and, the amount of initial rotation on the collapse dynamics and magnetic field 
amplification of a magnetized gas cloud. We note that there are two main time scales in the problem -
the free-fall time, $t_{\rm ff} \sim 1/\sqrt{G\,\rho}$ and the eddy turnover time scale, $t_{\rm ed} = l/v(l)$. 
For a $k^{-2}$ spectrum which we adopt here, $v(l) \sim l^{1/2}$ and therefore, $t_{\rm ed} \sim l^{1/2}$
on scales smaller than the injection scale. In paper II we showed that the driving scale of turbulence 
generated during gravitational collapse is of the order of the local Jeans length. \citet{KH10} show that 
it is the very process of formation of structures in the Universe on all scales that drives the turbulence. 
Given our initial core density, (subsection 2.1) and using the fact that turbulence is driven on the 
local Jeans length ($\lambda_{\rm J} \sim 1.5\,{\rm pc}$ at $\tau=0$) in our simulations, we have 
$t_{\rm ed} \sim 68.157\,{\rm yr}$ which is much smaller than the initial free-fall time scale 
$t_{\rm ff} = 0.56\,{\rm Myr}$. On a scale $\sim 0.1\,\lambda_{\rm J}$, the eddy turnover time scale is even 
smaller which demonstrates that significant dynamo action will occur on smaller scales provided
such scales are resolved. In addition to the above two time scales, in presence of rotation, there 
is an additional time scale $t_{\rm rot} \sim \Omega^{-1}$ where $\Omega$ is the angular velocity. 
For $\beta=8\%$, $t_{\rm rot} \sim 1.38\,{\rm Myr}$. 

\label{initcond}
\subsection{Effect of initial turbulence}
In this section, we explore the influence of varying the initial rms turbulent velocity of the cloud 
on the collapse and magnetic field amplification. The ubiquity of turbulence in the cloud core is 
central to the idea of magnetic amplification by small-scale dynamo action. While subsonic 
turbulence correspond to typical velocities found in the first star-forming minihalos, supersonic 
velocities may correspond to more massive systems like the first galaxies \citep{WTA08, Greif+08}.

\subsubsection{ \bf Evolution of the density and the velocity} 

\begin{figure*}
\centerline{\includegraphics[width=1.4\columnwidth]{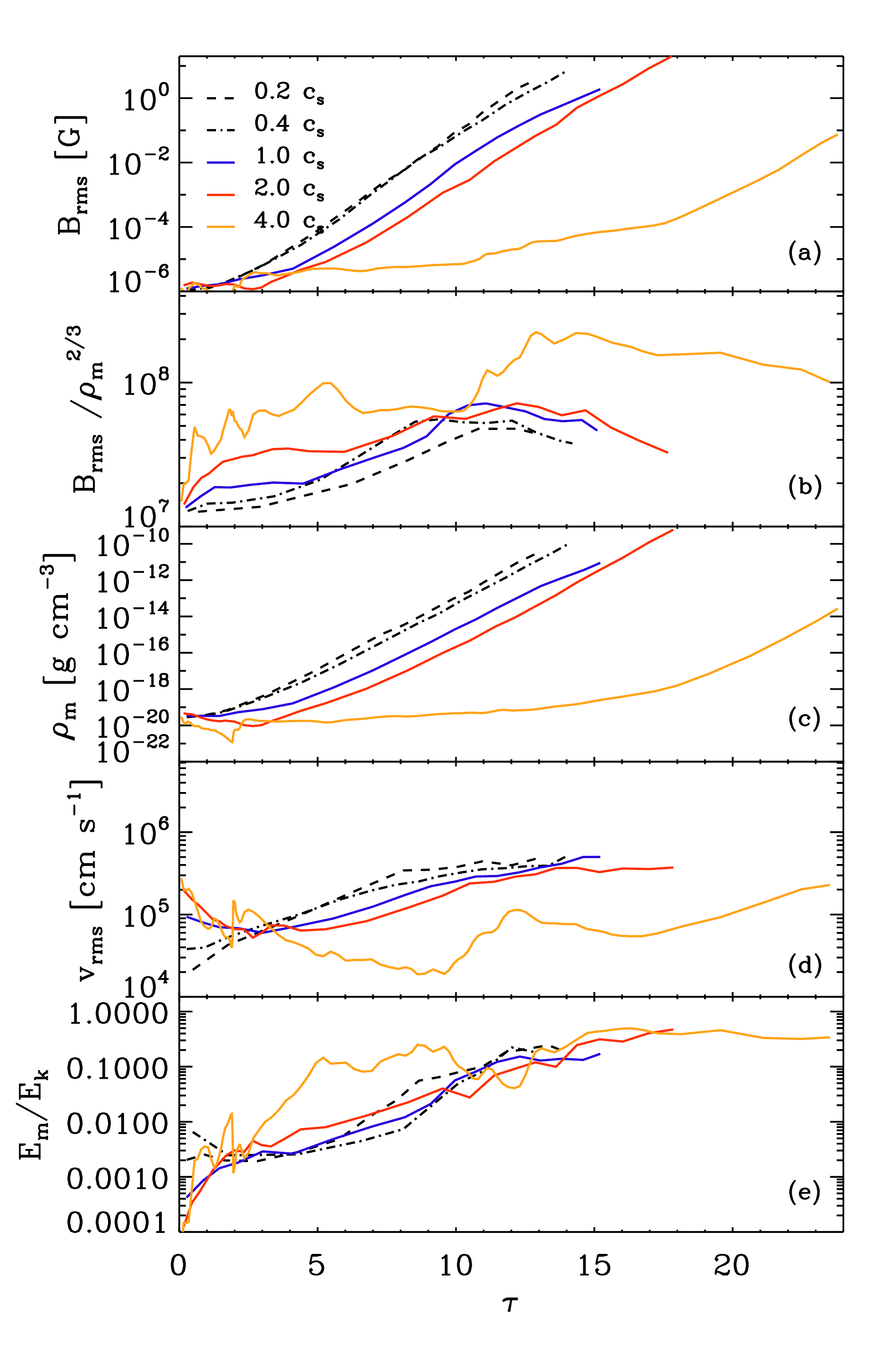}}
\caption{Evolution of the dynamical quantities in the central Jeans volume as a function of $\tau$ 
for runs with different initial Mach numbers of the turbulence: $v_{\rm rms}/c_{\rm s} = 0.2, 0.4, 1.0, 2.0, 4.0$. 
Panel (a) shows the rms magnetic field strength $B_{\rm rms}$, 
(b) the evolution of {\bf $B_{\rm rms}/\rho_{\rm m}^{2/3}$}, showing the turbulent dynamo amplification 
by dividing out the maximum possible amplification due to perfect flux freezing during spherical 
collapse, (c) the evolution of the mean density $\rho_{\rm m}$, (d) the rms velocity $v_{\rm rms}$, 
and (e) the ratio of magnetic to kinetic energy, $E_{\rm m}/E_{\rm k}$. All the simulations correspond to 
a resolution of 64 cells and do not include any initial rotation of the cloud. 
\label{tau_mach}}
\end{figure*}

In order to focus on the effects of turbulence, we do not include any ordered rotation of the primordial cloud. 
Figs.~\ref{tau_mach} and \ref{mdens_mach} show the plot of different physical quantities against the 
normalized time, $\tau$ and the mean density $\rho_{\rm m}$ respectively, for five different values of the 
initial rms turbulent velocity, 
$v_{\rm rms} = 0.2\,c_{\rm s}, 0.4\,c_{\rm s}, c_{\rm s}, 2\,c_{\rm s}$ and $4\,c_{\rm s}$.
The $\alpha$ parameter corresponding to these values are shown in Table~\ref{sumsim}. 
We first focus on the hydrodynamic aspects of these curves, namely the evolution of the mean density 
(panel c in Fig.~\ref{tau_mach}) and the rms velocity (panels d and c in Figs.~\ref{tau_mach} and 
\ref{mdens_mach} respectively). It turns out that initializing the collapse with different values of the rms 
turbulent velocity has an effect on how early or late runaway collapse sets in. This is partly due to the fact 
that the turbulent kinetic energy density prevents the gas from collapsing at the outset for initial transonic 
and supersonic velocities, while for subsonic velocities, the gas goes into collapse from $\tau \sim 1$. 
This is shown in the evolutionary behavior of the mean density ($\rho_{\rm m}$). The run 
with an initial $v_{\rm rms}\sim4\,c_{\rm s}$ does not undergo runaway collapse till about $\tau \sim 11 - 12$ 
because the initial $\alpha$ parameter ($\alpha = 3.72$) is high enough to prevent the runaway collapse 
till the time, the turbulence has decayed. The early stage of the evolution is marked by frequent changes in 
the rms velocity (panel (c) in Fig.~\ref{mdens_mach}) in the density range $(\sim 10^{-20} - 10^{-19})\,\rm{g}\,\,{\rm cm}^{-3}$. 
The evolution of the rms velocity for runs with initial $v_{\rm rms} = 1\,c_{\rm s}, 2\,c_{\rm s}$ is marked by a 
distinct decay phase till $\tau \sim 3$, after which both curves start to increase as the 
collapse regenerates the turbulence. 

\subsubsection{\bf Magnetic field evolution and signatures of dynamo saturation}

\begin{figure*}
\centerline{\includegraphics[width=1.4\columnwidth]{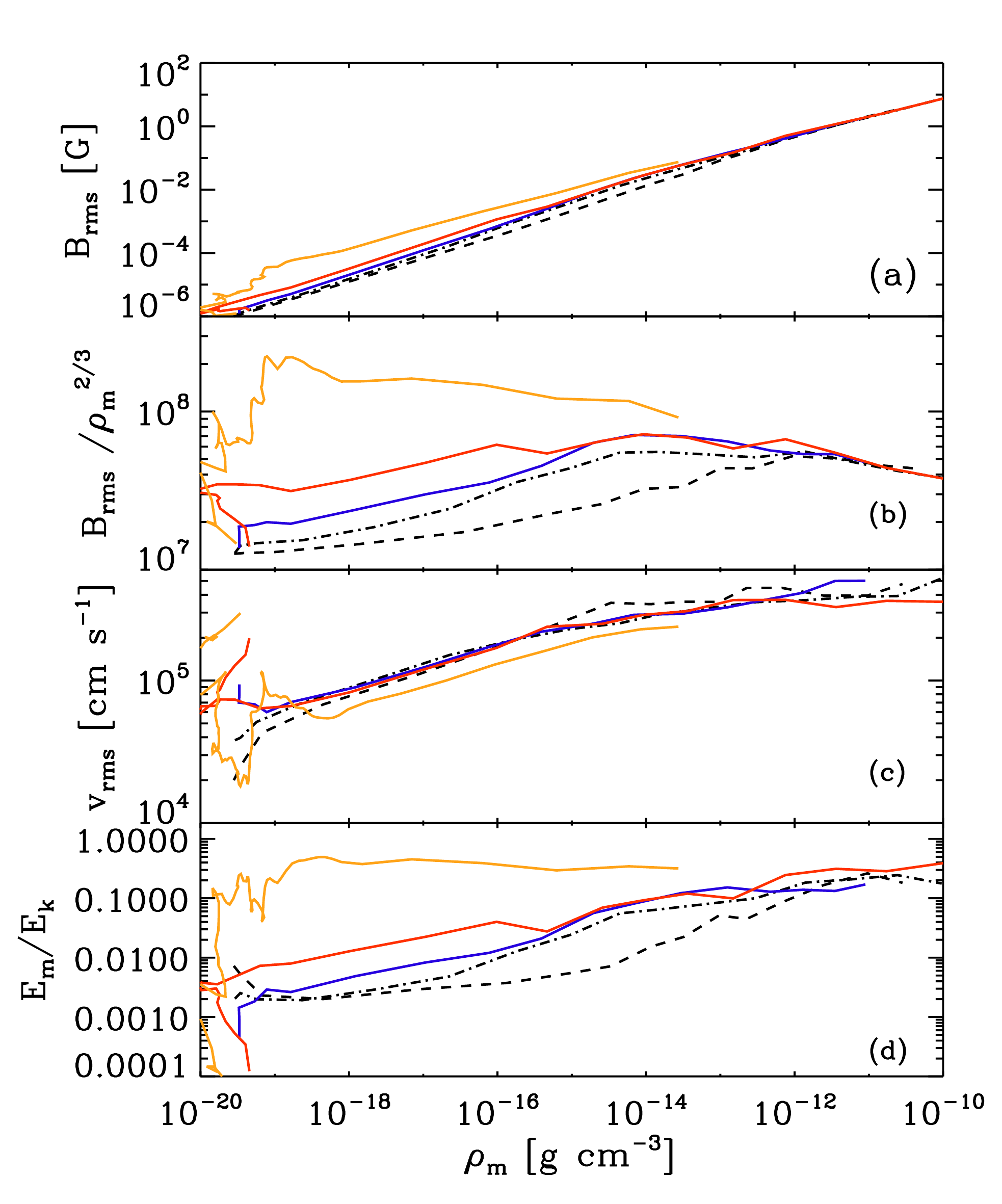}}
\caption{Same as Fig.~\ref{tau_mach}, but now plotted as a function of $\rho_{\rm m}$. 
\label{mdens_mach}}
\end{figure*}

\begin{figure*}
\centerline{\includegraphics[width=1.2\columnwidth]{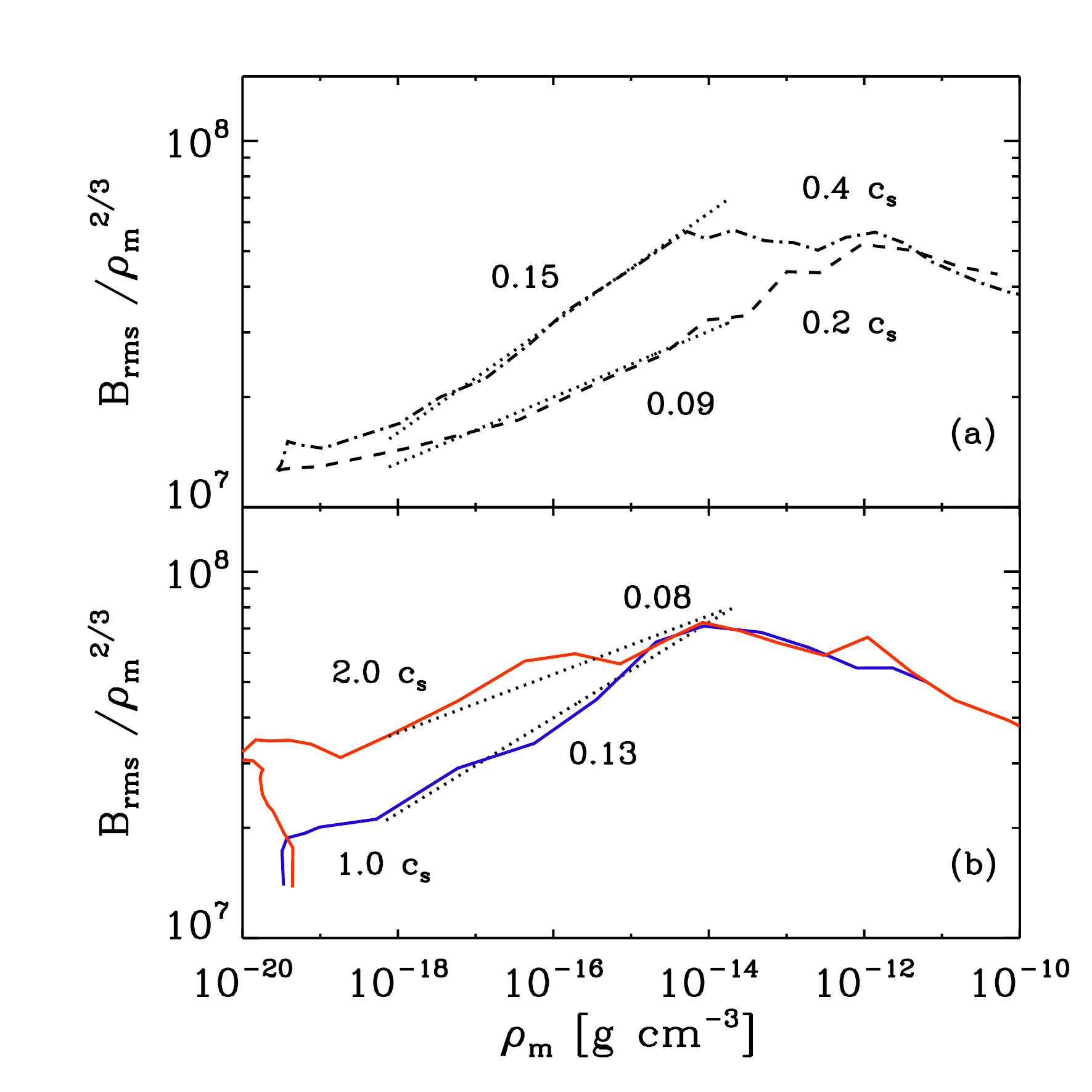}}
\caption{Growth rate of $B_{\rm rms}/\rho_{\rm m}^{2/3}$ for runs with $v_{\rm rms}= 0.2\,c_{\rm s}, 
0.4\,c_{\rm s}, 1.0\,c_{\rm s}$ and $2.0\,c_{\rm s}$. The growth rates for the different runs are 
computed in the density interval $\rho_{\rm m} \sim 10^{-18} - 10^{-14}\,\rm{g}\,\,{\rm cm}^{-3}$ when 
all the different runs are in identical stages of gravitational collapse. The growth rate of the dynamo 
generated magnetic field at first increases with increasing subsonic Mach numbers and then starts
to decrease in the supersonic regime.
\label{mach_mdens}}
\end{figure*}

The total magnetic field is amplified by more than 6 orders of magnitude, reaching peak values of 
$\sim 1\,{\rm G}$ at densities of $\sim 10^{-12}\,\rm{g}\,\,{\rm cm}^{-3}$ evident from panel (a) in 
Fig.~\ref{mdens_mach}. The evolution of the dynamo generated magnetic field is shown in panel (b)
of Figs.~\ref{tau_mach} and \ref{mdens_mach}. Except for the run with $v_{\rm rms} = 4\,c_{\rm s}$, 
the exponential amplification of the magnetic field by the small-scale dynamo continues till $\tau\sim 10$, 
illustrating the kinematic phase of the dynamo. This phase corresponds to densities up to 
$\approx 10^{-14}\,\rm{g}\,\,{\rm cm}^{-3}$ for runs with initial $v_{\rm rms} = (0.2 - 2.0)\,c_{\rm s}$. The 
exponential phase lasts till $\rho_{\rm m}\sim 10^{-13}\,\rm{g}\,\,{\rm cm}^{-3}$ for $v_{\rm rms} = 0.1\,c_{\rm s}$. 
The curves attain a peak value of $B_{\rm rms}/\rho_{\rm m}^{2/3}$ in the range $\sim (2 - 6) \times 10^{7}$. 
The behavior of the run with $v_{\rm rms} = 4\,c_{\rm s}$ is different. Here the small-scale dynamo 
field undergoes rapid fluctuations when the core density is in the range $(10^{-20} - 10^{-19})\,\rm{g}\,\,{\rm cm}^{-3}$ 
attaining a peak value of $2\times 10^{8}$ with respect to the initial value. This is an order of magnitude higher 
than the other runs. It is only after attaining core densities of $\sim 10^{-18}\,\rm{g}\,\,{\rm cm}^{-3}$ by when the 
rearrangement is complete, that it begins to resemble the other runs. In Fig.~\ref{mach_mdens}, we show the 
estimated growth rate of $B_{\rm rms}/\rho_{\rm m}^{2/3}$ computed in the density range 
$\rho_{\rm m} \sim (10^{-18} - 10^{-14})\,\rm{g}\,\,{\rm cm}^{-3}$
for runs with different initial Mach numbers. This approach ensures that the estimated growth rates correspond to 
the same stage of collapse for all the runs. The dynamo growth rate at first increases for subsonic Mach 
numbers and then decreases in the supersonic regime. Table~\ref{machsim} presents a summary 
of the estimated growth rates. Simulations of small-scale dynamo action by \citet{Federrath+11b} 
using external forcing also find a similar trend. They find a general increase of the growth rate with the 
Mach number, but a clear drop of the growth rate at the transition from subsonic to supersonic turbulence.
They attribute this drop in the growth rate to the formation of shocks at transonic speeds, which destroy
some of the coherent small-scale magnetic field structures necessary to drive dynamo amplification.  
The exponential phase is also marked by an increase in the ratio of the magnetic to kinetic energies, 
$E_{\rm m}/E_{\rm k}$ shown in panels (e) and (d) in Figs.~\ref{tau_mach} and \ref{mdens_mach}. 

In the density range $\rho_{\rm m}\sim(10^{-14} - 10^{-10})\,\rm{g}\,\,{\rm cm}^{-3}$, panel (b) in Fig.~\ref{mdens_mach}
shows a clear change in the slope of $B_{\rm rms}/\rho_{\rm m}^{2/3}$. As we argued in Section~\ref{Intro}, such a 
change of slope with $B_{\rm rms}/\rho_{\rm m}^{2/3}$ either remaining constant or decreasing is an indication of the 
fact that the dynamo proceeds to the saturation phase. The total field amplification in this phase is then only due to 
gravitational compression of the field lines. For an initial $v_{\rm rms} = 4.0\,c_{\rm s}$, the saturation phase begins 
as early as $\rho_{\rm m}\sim 10^{-19}\,\rm{g}\,\,{\rm cm}^{-3}$. Panels (e) and (d) in Figs.~\ref{tau_mach} and 
\ref{mdens_mach} respectively also show a gradual transition to saturation with values reaching about $0.2 - 0.4$. 
Comparing the evolution of $B_{\rm rms}/\rho_{\rm m}^{2/3}$ to that of $E_{\rm m}/E_{\rm k}$, we find the former to be 
better suitable as a saturation indicator for small-scale dynamo action in self-gravitating systems. While there is a distinct 
change in the slope of $B_{\rm rms}/\rho_{\rm m}^{2/3}$ after $\rho_{\rm m}=10^{-14}\,\rm{g}\,\,{\rm cm}^{-3}$ (panel b in 
Fig.~\ref{mdens_mach}), $E_{\rm m}/E_{\rm k}$ still continues to increase mildly attaining values in the range $0.2 - 0.4$. 
However, for the run with $v_{\rm rms} = 4.0\,c_{\rm s}$, both $B_{\rm rms}/\rho_{\rm m}^{2/3}$ and $E_{\rm m}/E_{\rm k}$
traces out the saturation phase equally well. A convincing probe of dynamo saturation is to explore the time evolution
of the spectra of the magnetic field. However, since the resolution of all the simulations discussed in this subsection 
is only 64 grid cells per Jeans length, we defer the discussion of the magnetic spectra to subsection 3.3. 
Nevertheless, the results obtained from our Mach number study demonstrate that our choice of the initial field strength 
allows us to probe two regimes of field amplification - an initial phase where both gravitational compression and the 
small-scale dynamo amplify the field followed by a phase of saturation of the dynamo in which only compression drives 
the field amplification.   

%%%%%%%%%%%%%%%%%%%%%%%%%%%%%%%%%%%%%%%%%%%%%%%%%%%%
\begin{table}
%\begin{minipage}{126mm}
\begin{center}
\begin{tabular}{ccc} \hline \hline 
Simulation Run & Mach Number $[v_{\rm rms}/c_{\rm s}]$ & Growth rate\\ \hline \hline
R64M0.2rot0 & 0.2 & 0.09\\ \hline 
R64M0.4rot0 & 0.4 & 0.15 \\ \hline
R64M1.0rot0 & 1.0 & 0.13\\ \hline
R64M2.0rot0 & 2.0 & 0.08 \\ \hline \hline
\end{tabular}
\end{center}
\caption{Comparison of the growth rates of $B_{\rm rms}/\rho_{\rm m}^{2/3}$ as a function of the mean 
density for simulations with different initial Mach numbers taken from figure 3. The growth rates are computed 
when the different runs are in the same stage of the collapse. The table shows that the growth rate of the dynamo 
generated field at first increases with the increase in initial subsonic Mach number but then starts decreasing 
for supersonic Mach numbers.}
\label{machsim}
%\end{minipage}
\end{table}
%%%%%%%%%%%%%%%%%%%%%%%%%%%%%%%%%%%%%%%%%%%%%%%%%%%
\subsection{Effect of the injection scale}
In this subsection we probe the effect of different injection scales of the initial turbulence on the 
gravitational collapse and magnetic field amplification by the small-scale dynamo. We consider 
two simulations with the injection scale peaked on scales $l_{\rm inj}/\lambda_{\rm J} = 0.17$ 
and $l_{\rm inj}/\lambda_{\rm J} = 0.7$. As before, we do not include any ordered rotation for these runs and 
use a resolution of $64$ cells to resolve the local Jeans length. Both the simulations start with the same value 
of the transonic velocity dispersion. 

\begin{figure*}
\centerline{\includegraphics[width=1.4\columnwidth]{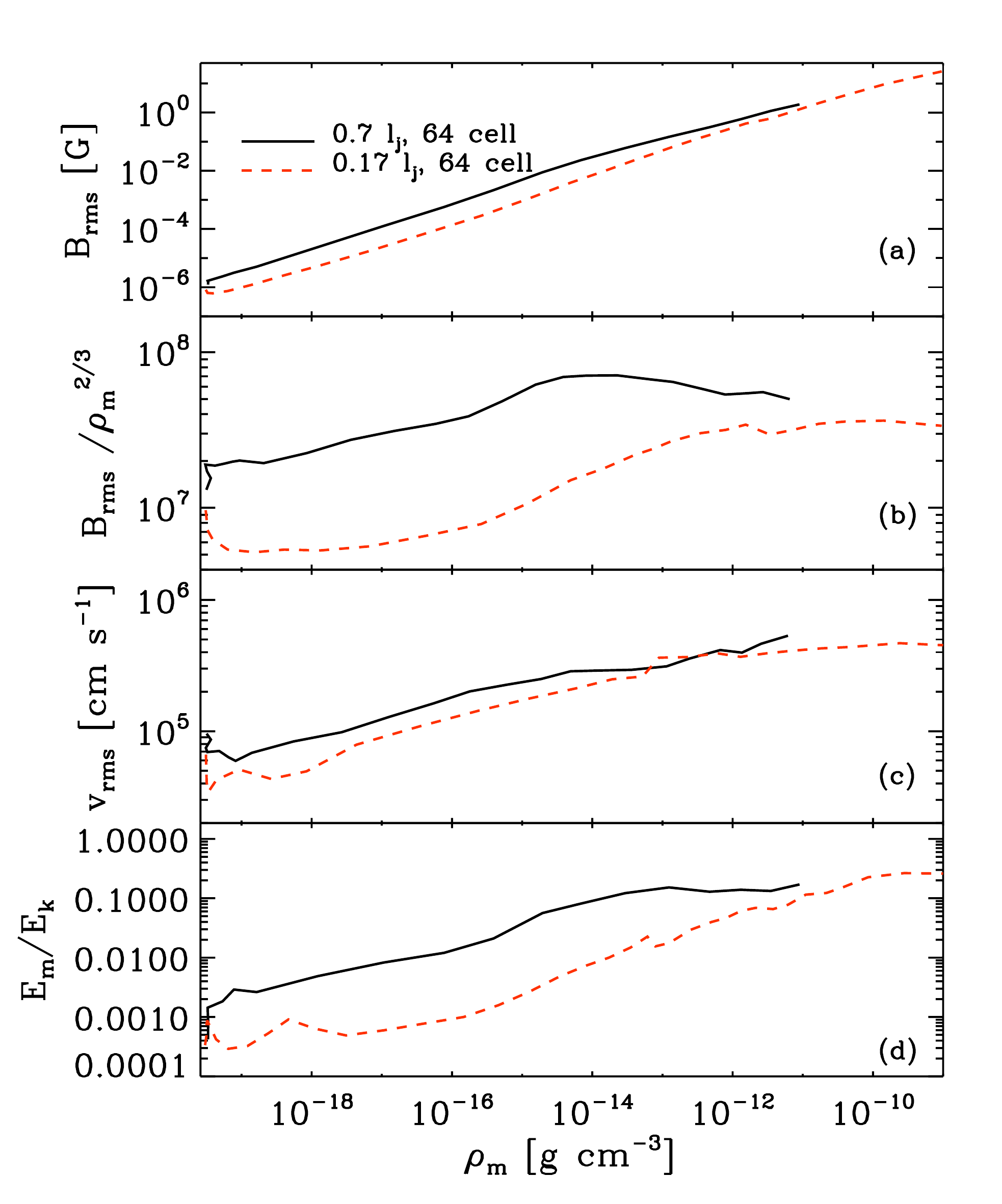}}
\caption{Same as in Fig.~\ref{mdens_mach}, but here we compare two runs with zero initial rotation, initial 
turbulence with $v_{\rm rms}/c_{\rm s} = 1.0$ and Jeans resolution of 64 cells for different initial injection 
scales of turbulence: 0.7\,$l_{\rm j}$ (black solid line) and 0.17\,$l_{\rm j}$ (red dashed line). When initial 
turbulence is injected on a smaller scale, the turbulence decays faster leading to an initial decay of 
$B_{\rm rms}/\rho_{\rm m}^{2/3}$. 
\label{allcomb_lj_mdens}}
\end{figure*}

Figure~\ref{allcomb_lj_mdens} shows the variation of the various physical quantities as a function of
the mean density, $\rho_{\rm m}$. With a smaller injection scale, the turbulence decays faster 
(panel (c) in Fig.~\ref{allcomb_lj_mdens}) compared to the run where $l_{\rm inj} = 0.7\,\lambda_{\rm J}$ as 
more kinetic energy has to be initialised on the smaller scales to get the same overall turbulent Mach 
number. This can also be seen by comparing the density snapshots in the first column of Fig.~\ref{lj_slices}
where the arrows denote the velocity vectors. These correspond to times when the central core 
density in both the runs are $\sim 10^{-17}\,\rm{g}\,\,{\rm cm}^{-3}$. The upper plot in this column 
corresponds to a simulation with $l_{\rm inj}/\lambda_{\rm J} = 0.17$ while the plot on the lower panel 
corresponds to $l_{\rm inj}/\lambda_{\rm J} = 0.7$. The run with a smaller injection scale shows less 
turbulent motions inside the Jeans volume compared to the run where the initial injection scale is peaked 
on scales of the order $0.7\,\lambda_{\rm J}$. 

The second column in Fig.~\ref{lj_slices} shows that the total magnetic field attains a peak value of 
$\sim 10^{-4}\,{\rm G}$ for $l_{\rm inj}/\lambda_{\rm J} = 0.17$ and about $\sim 10^{-3}\,{\rm G}$ for the run 
with $l_{\rm inj}/\lambda_{\rm J} = 0.7$ at $\rho_{\rm m} = 10^{-17}\,\rm{g}\,\,{\rm cm}^{-3}$. The faster 
decay in the initial turbulence for the smaller injection scale leads to the decay in $B_{\rm rms}/\rho_{\rm m}^{2/3}$ 
as evident from panel (b) in Fig.~\ref{allcomb_lj_mdens}. 
Over time, as the turbulence gets regenerated during the collapse, the magnetic field and hence the 
magnetic energy gets amplified by the dynamo process. The kinematic phase of the dynamo extends 
to $\rho_{\rm m}\sim 10^{-12}\,\rm{g}\,\,{\rm cm}^{-3}$ for $l_{\rm inj} = 0.17\,\lambda_{\rm J}$ with a peak 
value of $B_{\rm rms}/\rho_{\rm m}^{2/3}\sim 2.5\times 10^{7}$ and up to $\sim 10^{-15}\,\rm{g}\,\,{\rm cm}^{-3}$ 
for $l_{\rm inj} = 0.7\,\lambda_{\rm J}$ with a peak value of $\sim 7\times 10^{7}$. Saturation of the 
dynamo is once again clearly illustrated by the change in slope of $B_{\rm rms}/\rho_{\rm m}^{2/3}$. 
For $l_{\rm inj} = 0.17\,\lambda_{\rm J}$, this occurs from a density $\sim 10^{-10}\,\rm{g}\,\,{\rm cm}^{-3}$
while for $l_{\rm inj} = 0.7\,\lambda_{\rm J}$, saturation starts from $\rho_{\rm m}\sim 10^{-14}\,\rm{g}\,\,{\rm cm}^{-3}$. 
Comparing this with the evolution of $E_{\rm m}/E_{\rm k}$ in panel (d), we find that the ratio of the 
energies still continues to increase up to $\rho_{\rm m}\sim 10^{-10}\,\rm{g}\,\,{\rm cm}^{-3}$ for 
$l_{\rm inj} = 0.17\,\lambda_{\rm J}$ and up to $\sim 10^{-13}\,\rm{g}\,\,{\rm cm}^{-3}$ for 
$l_{\rm inj} = 0.7\,\lambda_{\rm J}$. The magnetic energy saturates at about $0.1$ of the equipartition value 
for this case while for a smaller injection scale, the saturation level is at $\sim 0.3$ of the equipartition value.

\subsection{Effect of initial uniform rotation}

Finally, we explore the effects of rotation on the collapse and magnetic field amplification. We begin by 
comparing simulations performed at the same resolution (128 cells per Jeans length) with the initial rotation 
parameter $\beta$ having values of $0\%$, $4\%$ and $8\%$. 

\subsubsection{\bf Evolution of the density and the velocity}

\begin{figure*}
\centerline{\includegraphics[width=1.4\columnwidth]{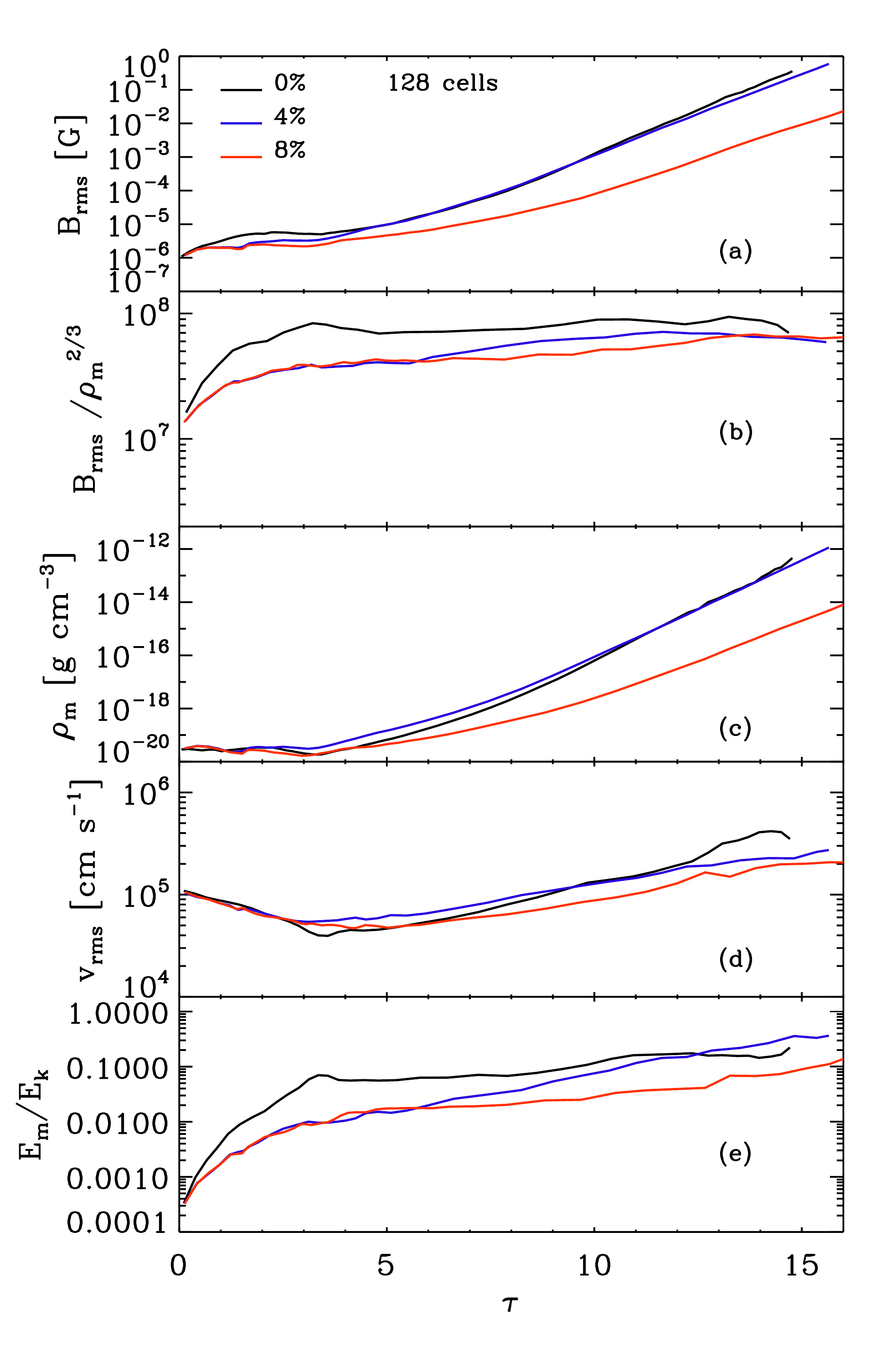}}
\caption{Evolution of the dynamical quantities in the central Jeans volume as a function of $\tau$ 
for runs with different initial rotation. Panel (a) shows the rms magnetic field strength $B_{\rm rms}$, 
(b) the evolution of {\bf $B_{\rm rms}/\rho_{\rm m}^{2/3}$}, showing the turbulent dynamo amplification 
by dividing out the maximum possible amplification due to perfect flux freezing during spherical 
collapse, (c) the evolution of the mean density $\rho_{\rm m}$, (d) the rms velocity $v_{\rm rms}$, 
and (e) the ratio of magnetic to kinetic energy, $E_{\rm m}/E_{\rm k}$. All the simulations correspond to 
a resolution of 128 cells per Jeans length.
\label{allcomb_rot}}
\end{figure*}

In Fig.~\ref{allcomb_rot} we show the evolution of various physical quantities as a function of $\tau$ for 
runs with $\beta = 0\%, 4\%$ and $8\%$. The same physical quantities are plotted as a function of 
$\rho_{\rm m}$ in Fig.~\ref{allcomb_rot_mdens}. As reported earlier in Papers I and II, the dynamical 
evolution of the system shows two distinct phases. First, as the initial turbulent velocity decays, the system 
exhibits weak oscillatory behavior (up to a time $\tau=4$) with the mean density $\rho_{\rm m}$ evolving 
similarly for runs with $0\%$, $4\%$ and $8\%$ initial uniform rotation. This is evident from panels 
(c) and (d) in Figs.~\ref{allcomb_rot_mdens} and \ref{allcomb_rot} respectively. After this, runaway collapse 
sets in with peak densities of the order of $\sim 10^{-12}\,\rm{g}\,\,{\rm cm}^{-3}$ being attained for $\beta = 0\%$
and $4\%$. Higher value of the initial rotation (e.g. $\beta = 8\%$) leads to a delayed collapse as the cloud 
is more rotationally supported compared to the other two cases (panel (c) in Fig.~\ref{allcomb_rot}). After 
the initial decay, the rms velocity increases as the turbulence is regenerated by the collapse.

\subsubsection{\bf Magnetic field evolution}

\begin{figure*}
\centerline{\includegraphics[width=1.6\columnwidth]{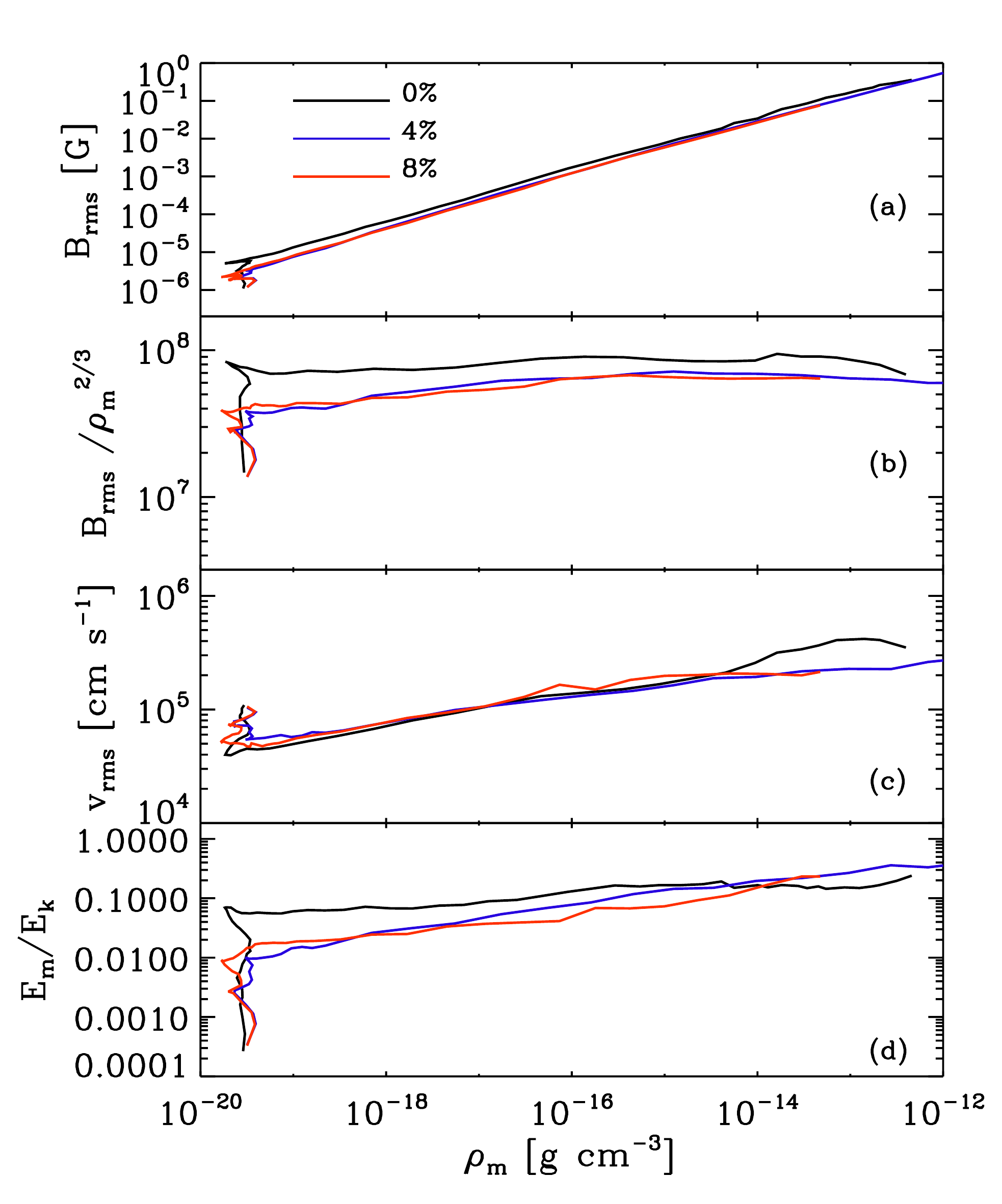}}
\caption{Same as Fig.~\ref{allcomb_rot}, but now plotted versus the mean density.
\label{allcomb_rot_mdens}}
\end{figure*}

Comparing the evolution of the rms magnetic field in the three runs we find from panel (a) in 
Fig.~\ref{allcomb_rot_mdens}, the magnetic field increases by more than 5 orders of magnitude starting 
from $1\,\mkG$ reaching strengths of $0.1\,{\rm G}$ at $\rho_{\rm m} \sim 10^{-13}\,\rm{g}\,\,{\rm cm}^{-3}$. 
The corresponding evolution with respect to $\tau$ shown in panel (a) of Fig.~\ref{allcomb_rot} may seem 
to suggest that the field amplification is weaker for the $8\%$ run compared to the other two. Panel (b) in 
both Figs.~\ref{allcomb_rot} and \ref{allcomb_rot_mdens} shows the evolution of the magnetic field arising 
due to turbulent motions, i.e, due to small-scale dynamo action. For all the three different runs, the field is 
exponentially amplified in the initial phase of decaying turbulence (till about $\tau=3 - 4$). After this, the 
dynamo continues to amplify the field up to $\tau \sim 13$ or $\rho_{\rm m}\sim 10^{-16}\,\rm{g}\,\,{\rm cm}^{-3}$ 
but the amplification is no longer exponential. The ratio of the magnetic to kinetic energies (panels e and d 
in Figs.~\ref{allcomb_rot} and \ref{allcomb_rot_mdens} also show similar behavior with an initial exponential 
increase followed by a phase of linear growth. 

\subsubsection{\bf Dynamo saturation}

\begin{figure*}
\centerline{\includegraphics[width=1.8\columnwidth]{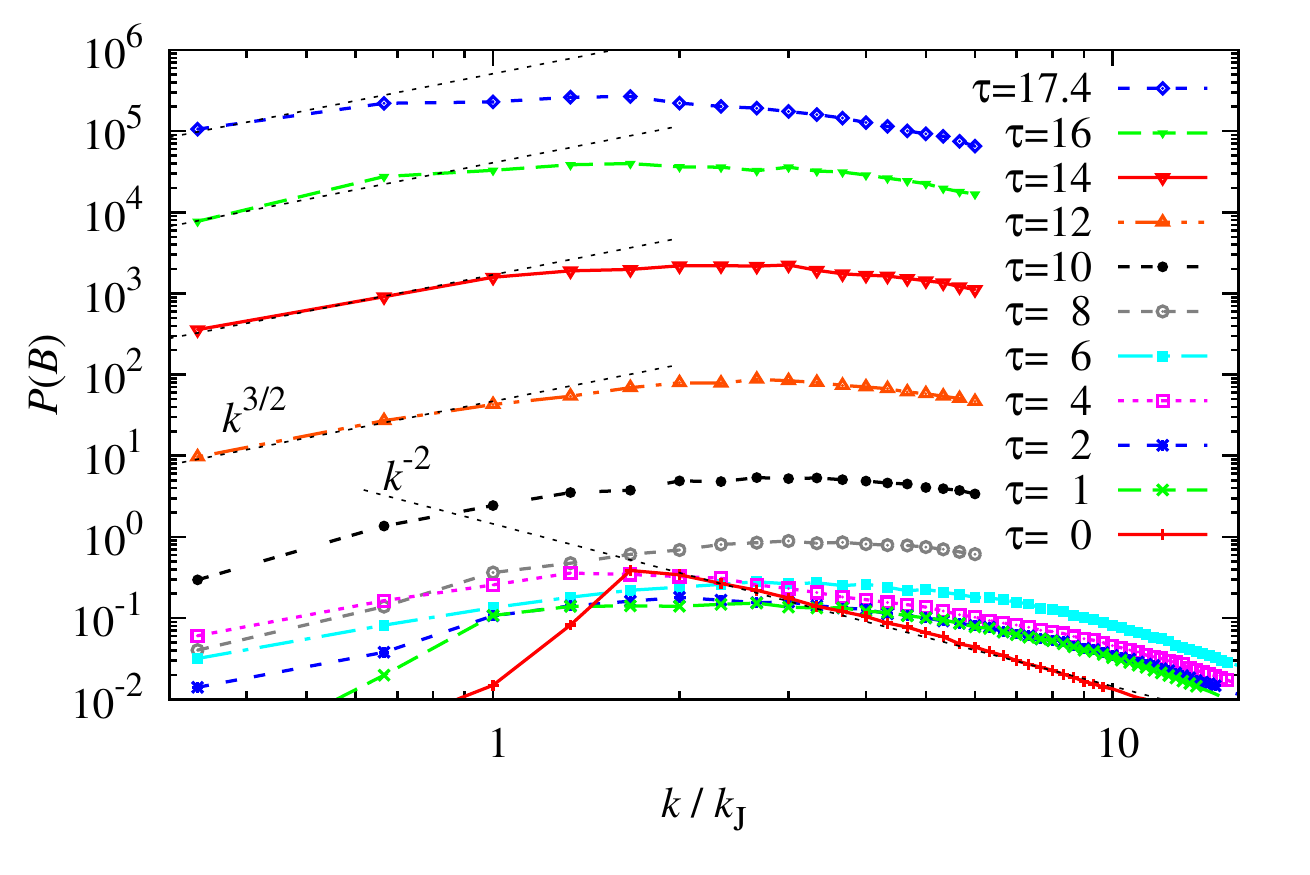}}
\caption{Time evolution of the magnetic field spectra for the run with 128 cells per Jeans length and 
$E_{\rm rot}/|E_{\rm grav}| = 8\%$. The spectra are plotted against the wavenumber normalised to the local 
Jeans wavenumber. The initial magnetic field spectrum ($\tau = 0$) follows a $k^{-2}$ power-law on scales 
smaller than the peak scale, $k/k_{\rm J}\approx 1.4$ as determined by the initial conditions. After the initial 
decay of the magnetic energy, the peak of the spectra shifts to $k/k_{\rm J} \approx 3-4$ and stays constant 
in this range up to $\tau = 12$. Because of the smaller eddy turnover times, these high $k/k_{\rm J}$ modes 
(smaller scales) attain saturation first. The field in the $k/k_{\rm J} < 3-4$ modes attain saturation beyond 
$\tau=12$ with the peak of the spectra now shifted to $k/k_{\rm J} \approx 1$ by the end of the simulation. 
In addition to the amplification by the dynamo, the magnetic field is also simultaneously amplified by gravitational 
compression. This leads to an overall rise of the spectrum with time. 
\label{mag_spec}}
\end{figure*}

Saturation of the dynamo occurs from $\rho_{\rm m}\sim 10^{-15}\,\rm{g}\,\,{\rm cm}^{-3}$ when all the three 
curves of $B_{\rm rms}/\rho_{\rm m}^{2/3}$ shows a change in slope. From panel (b) in Fig.~\ref{allcomb_rot_mdens}, 
these curves tend to remain almost constant (red curve) or decay slightly (black and blue curves). However, the 
$E_{\rm m}/E_{\rm k}$ ratio plotted in panel (d) continues to increase with a weak dependence on the mean density
eventually saturating at values of $0.2 - 0.4$. This once again highlights the fact that $B_{\rm rms}/\rho_{\rm m}^{2/3}$ 
is a better indicator of dynamo saturation in gravitating systems. A more detailed analysis of dynamo saturation 
can be obtained from a Fourier analysis of the spectra of the magnetic field. We recall that in Kolmogorov turbulence, 
the eddy turnover time $t_{\rm ed} = l/v \sim l^{2/3}$. Thus, smaller scale eddies amplify the magnetic field faster due 
to their shorter eddy turnover times. Therefore, saturation of the magnetic field should first occur on the smaller 
scales and then gradually be attained on the larger scales. 
To illustrate this phenomena in a more quantitative fashion, we perform a Fourier analysis  of the 
magnetic field spectra in the collapsing frame of reference for one of our rotation runs with $\beta = 8\%$. To 
compute the spectra, we extract the AMR data in a cube about three times the size of the local Jeans length. For 
more details on the data extraction procedure and the Fourier analysis we refer the reader to subsection 2.4 of Paper II. 
Figure~\ref{mag_spec} shows the time evolution of the magnetic energy spectrum as a function of the wavenumber, 
$k/k_{\rm J}$, i.e., normalized to the local Jeans wavenumber. Consistent with our initial conditions, the spectrum 
at $\tau=0$ shows the $P(B) \propto k^{-2}$ power-law scaling. In addition, the initial spectrum is peaked at 
$k/k_{\rm J}\approx 1.4$ which is in good agreement with our initial conditions where the initial Jeans length of the 
core is $1.5\,{\rm pc}$ and the peak of magnetic power spectrum is at $0.8\,{\rm pc}$. The time evolution of the 
magnetic spectrum shows two important features. Firstly, the peak of the initial magnetic spectrum quickly shifts to 
smaller scales starting from the initial $k/k_{\rm J}=1.4$ and then stays roughly constant at $k/k_{\rm J} \approx 3 - 4$ 
till $\tau = 12$. This corresponds to about 43 - 32 grid cells which is consistent with the Jeans resolution criterion 
proposed in Paper II. Next, from $\tau = 14$, the peak of the spectrum shifts to smaller $k/k_{\rm J}$ values, i.e., to
larger scales. The shift in the peak of the spectrum from $k/k_{\rm J} \approx 3 - 4$ to $k/k_{\rm J}\approx 1$ at late 
times implies that the magnetic field at first saturates on the smaller scales followed by saturation on larger scales. 
Thus, the information obtained from the time evolution of the magnetic field spectra is consistent with theoretical 
predictions \citep{S99, BS05}. The magnetic energy however continues to grow with time because the field continues 
to be amplified by the gravitational collapse.  

\begin{figure*}
\centerline{\includegraphics[width=1.8\columnwidth]{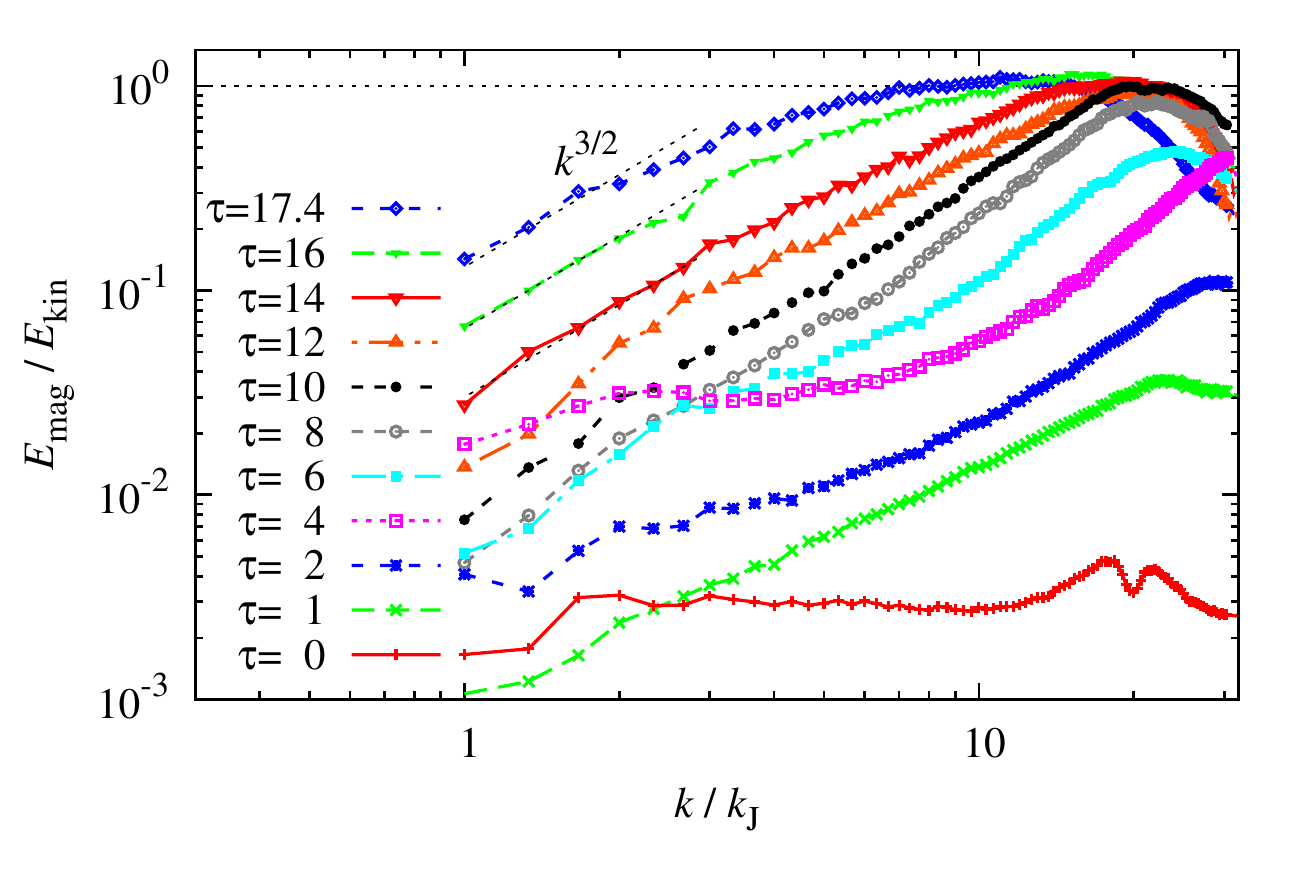}}
\caption{Time evolution of the spectra of the ratio of the magnetic to kinetic energies for the same simulation as 
in figure 8. As smaller scale eddies amplify the magnetic field faster, the peak of the spectrum initially shifts to smaller 
scales (i.e., larger $k/k_{\rm J}$ values). By $\tau\sim10-12$, the magnetic energy attains equipartition with the 
kinetic energy on scales $k/k_{\rm J}\sim 20$. Further time evolution shows that the peak now spreads to 
larger scales (i.e., smaller $k/k_{\rm J}$ values) with the field attaining equipartition on scales $k/k_{\rm J}=7-10$
at $\tau=17.4$. This plot therefore illustrates the saturation mechanism of the dynamo in a self-gravitating system. 
\label{emekin_spec}}
\end{figure*}

To further analyse the level of saturation we plot the time evolution spectra of the ratio of the magnetic to kinetic 
energies in Fig.~\ref{emekin_spec}. 
Since we extract about three times the volume for our Fourier analysis, we have taken a mean density of 
$\rho_{\rm m}\sim 10^{-21}\,\rm{g}\,\,{\rm cm}^{-3}$ resulting in an initial value of $E_{\rm m}/E_{\rm k}\sim 2\times10^{-3}$. 
We also remove all large-scale velocity contributions (i.e., global rotation and infall) from the kinetic energy. 
Since the smaller scale eddies amplify the field faster, $E_{\rm m}/E_{\rm k}$ first starts to peak at large $k/k_{\rm J}$ 
for $\tau > 0$. At $\tau=12$, the magnetic energy attains equipartition with the kinetic energy on a scale $k/k_{\rm J} = 20$. 
Once saturation has been attained on this scale, the peak of the spectrum now starts to shift to smaller $k/k_{\rm J}$. 
By $\tau=17.4$, the peak of the spectrum occurs at $k/k_{\rm J} = 7 - 10$, i.e., saturation of the dynamo is now achieved
on larger scales. This gradual shift of the peak of $E_{\rm m}/E_{\rm k}\sim 1$ to smaller $k/k_{\rm J}$ demonstrates 
the saturation of the dynamo and a possible development of a large-scale coherent magnetic field if the simulation is 
evolved further in time. The spectra further show that the magnetic field becomes dynamically important to back react
on the flow on scales $k/k_{\rm J} = 20$ by $\tau=12$ with the magnetic energy attaining equipartition with the kinetic 
energy. By the end of the simulation, equipartition is achieved on even larger scales.

\subsubsection{\bf Radial profiles} 

\begin{figure*}
\centerline{\includegraphics[width=1.12\columnwidth]{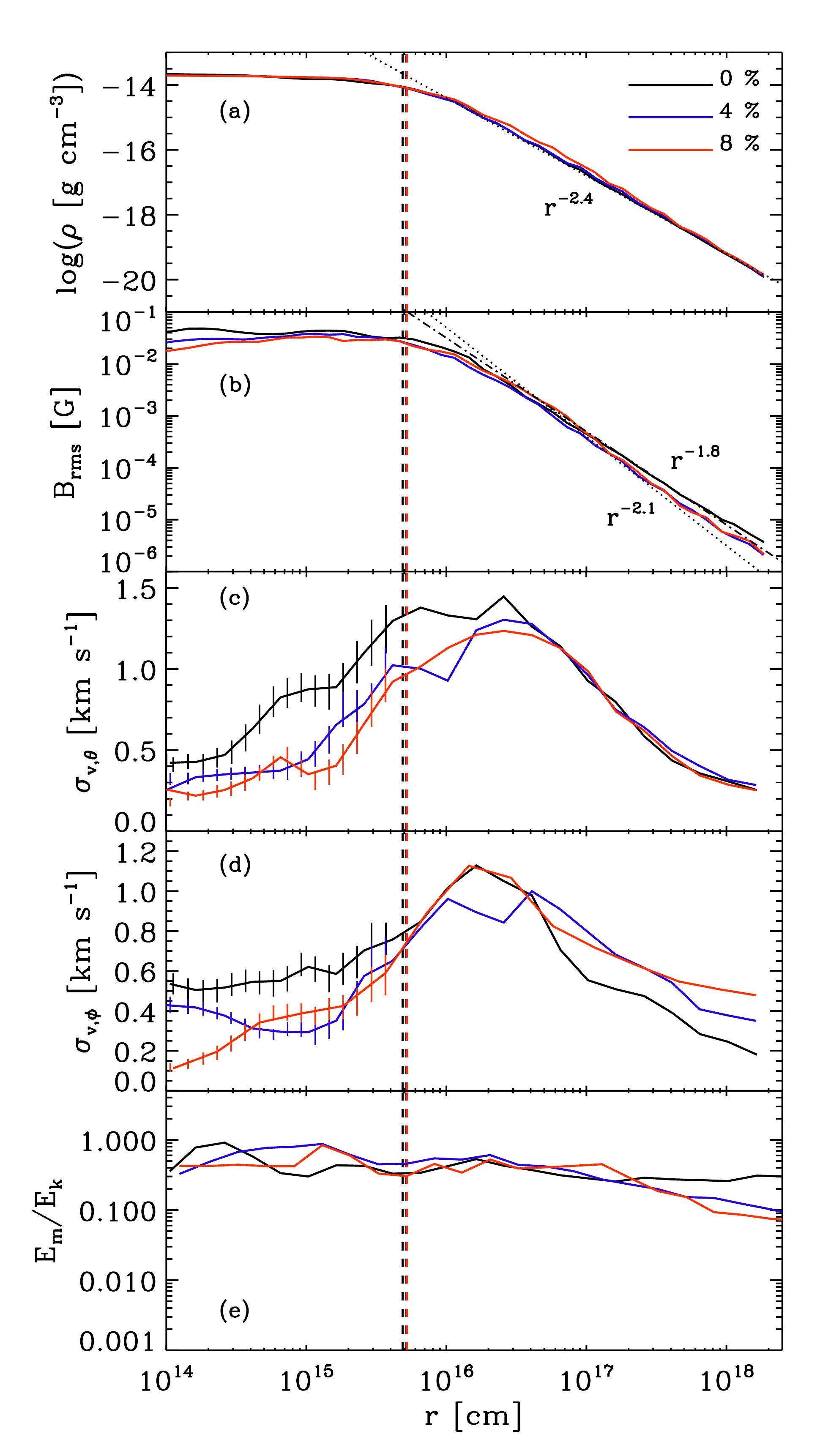}}
\caption{Radial profiles of the density in panel (a), the rms magnetic field in panel (b), 
components of the velocity dispersion, $\sigma_{\rm v,\theta}$ and $\sigma_{\rm v,\phi}$ 
in panels (c) and (d) respectively and the ratio of magnetic to kinetic energies in panel (e) 
at a time when runs with different initial rotation reached the same core density as shown 
in panel (a). Dashed vertical lines depict the local Jeans radius. Error bars are shown as solid 
vertical lines (panels c and d) inside the Jeans radius. The data are obtained from simulations 
with $\lambda_{\rm J}$ resolved by 128 cells. 
\label{velsigma_comb}}
\end{figure*}

Figure~\ref{velsigma_comb} shows the radial profile of the density, the rms magnetic field, the 
polar and azimuthal velocity dispersions, $\sigma_{\rm v, \theta}$ , $\sigma_{\rm v,\phi}$ and 
the ratio of magnetic to kinetic energies ($E_{\rm m}/E_{\rm k}$) for runs with $\beta=0, 4$ and $8\%$ 
initial rotation at a time when all the three attain the same core density. Similar to earlier results in 
Papers I and II, the density develops a flat inner core and falls off as $r^{-2.4}$ due to the effective 
equation of state with $\Gamma=1.1$ \citep[see][]{Larson69}. Panel (b) shows 
that the rms magnetic field attains peak values between $0.02 - 0.04\,{\rm G}$ within the Jeans 
volume. The radial profile of $B_{\rm rms}$ shows a radial dependence $\propto (r^{-1.8} - r^{-2.1})$
for the three runs. This is significantly steeper than the expectation from pure flux freezing where 
$B(r) \propto r^{-4/3}$. The velocity dispersion shown in panels (c) and (d) increases in the envelope 
and drops inside the Jeans volume. This is most likely due to the back reaction of the strong magnetic 
fields generated inside the Jeans volume. The ratio of the magnetic to kinetic energies plotted in 
panel (e) attains values in the range $\sim 0.3 - 0.8$ inside the Jeans volume.

\subsubsection{\bf Morphological features}

What effect does rotation have on the morphology of the cloud and does it lead to any change 
in the morphology of the small-scale dynamo generated field? In Fig.~\ref{dens_slices} we show 
two-dimensional snapshots of the density field in two planes: $x-y$ and $x-z$ for the run with 
$\beta=8\%$ at a time when the central core density is $\sim 10^{-15}\,\rm{g}\,\,{\rm cm}^{-3}$. 
The upper panel shows the zoomed-in slices, while the lower panel shows the zoomed-out version 
of the density field. Turbulent motions are dominant inside the Jeans volume as can be seen from 
the plotted velocity vectors in the zoomed-in slice in the $x-z$ plane. Rotational motion of the gas 
cloud is clearly seen from the $x-y$ slice in the lower panel. In conformity with known results, uniform 
rotation leads to a flattening of the gas cloud shown in the $x-z$ slice illustration in the lower panel. 

\begin{figure*}
\centerline{\includegraphics[width=2.0\columnwidth]{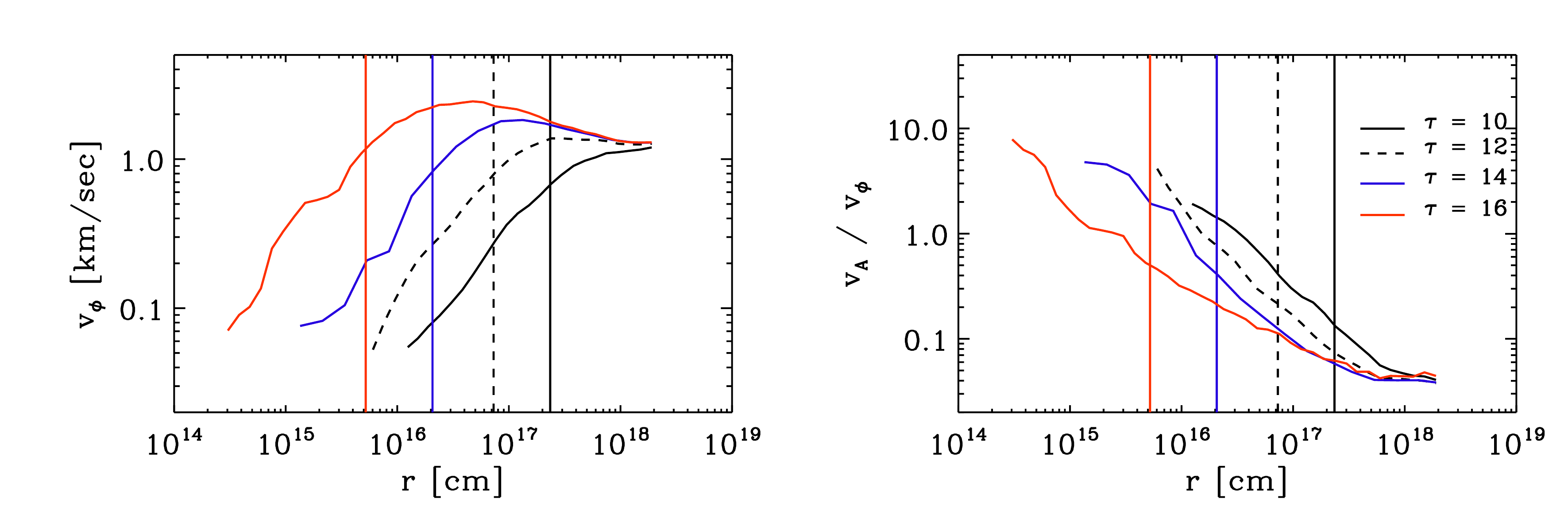}}
\caption{Radial profiles of the toroidal velocity $v_\phi$ (left panel) and the Alf\'ven velocity $v_{\rm A}$ 
normalised to $v_\phi$ (right panel) at different times for a run with $8\%$ initial rotation. The vertical lines 
indicate the local Jeans radius at different times. Runs plotted here are from simulations 
where $\lambda_{\rm J}$ is resolved with 128 cells. 
\label{vavphi}}
\end{figure*}

Quite interestingly, the morphology of the magnetic field in the saturated state does not seem to 
depend significantly on the amount of initial rotation (see Fig.~\ref{totmag_slices}). The magnetic field 
is randomly oriented within the local Jeans volume. This is reminiscent of the typical magnetic field 
structures seen in simulations of small-scale dynamo action where turbulence is driven artificially with 
random forcing \citep{BS05, Federrath+11b}.  In our case, the turbulence is driven completely by the 
gravitational collapse \citep[Papers I and II]{KH10}. In Fig.~\ref{totmag_slices} we plot the slices of the 
total magnetic field in $x-y$ and $x-z$ planes for two different cases: the first row corresponds to no net 
rotation and the second row corresponds to $8\%$ initial rotation. In both cases, the magnetic field is 
randomly oriented, attaining peak values of $0.1\,\mkG$ within the central Jeans volume. In Fig.~\ref{vavphi}, 
we show the radial profiles of the toroidal velocity $v_{\phi}$ (left panel) and the ratio of the Alf\'ven 
velocity normalized to the toroidal velocity (right panel) at different times. The toroidal velocity increases 
initially as we move from the outer to inner radii and then drops inside the local Jeans radius. This is because, 
inside the Jeans volume, a protostellar core of uniform density (see panel (a) in Fig.~\ref{velsigma_comb} 
where the density forms a flat inner core inside the Jeans radius) has formed which rotates like a solid 
body. The reason why uniform rotation does not introduce any morphological change in the 
magnetic field can be explained with the help of the radial profile of the ratio of the Alf\'ven to the rotational 
velocity shown in the above figure. At any given time, the increase in the ratio $v_{\rm A}/v_\phi$ as one 
proceeds from the outer to inner regions implies that the magnetic field is amplified on a much shorter 
timescale than the orbital time of the cloud. The dynamical effect of rotation on the magnetic field morphology 
will become observable at much later times, when the orbital timescale and the timescale on which the 
magnetic field is amplified become comparable.

\subsubsection{\bf Resolution effects}
\begin{figure*}
\centerline{\includegraphics[width=1.6\columnwidth]{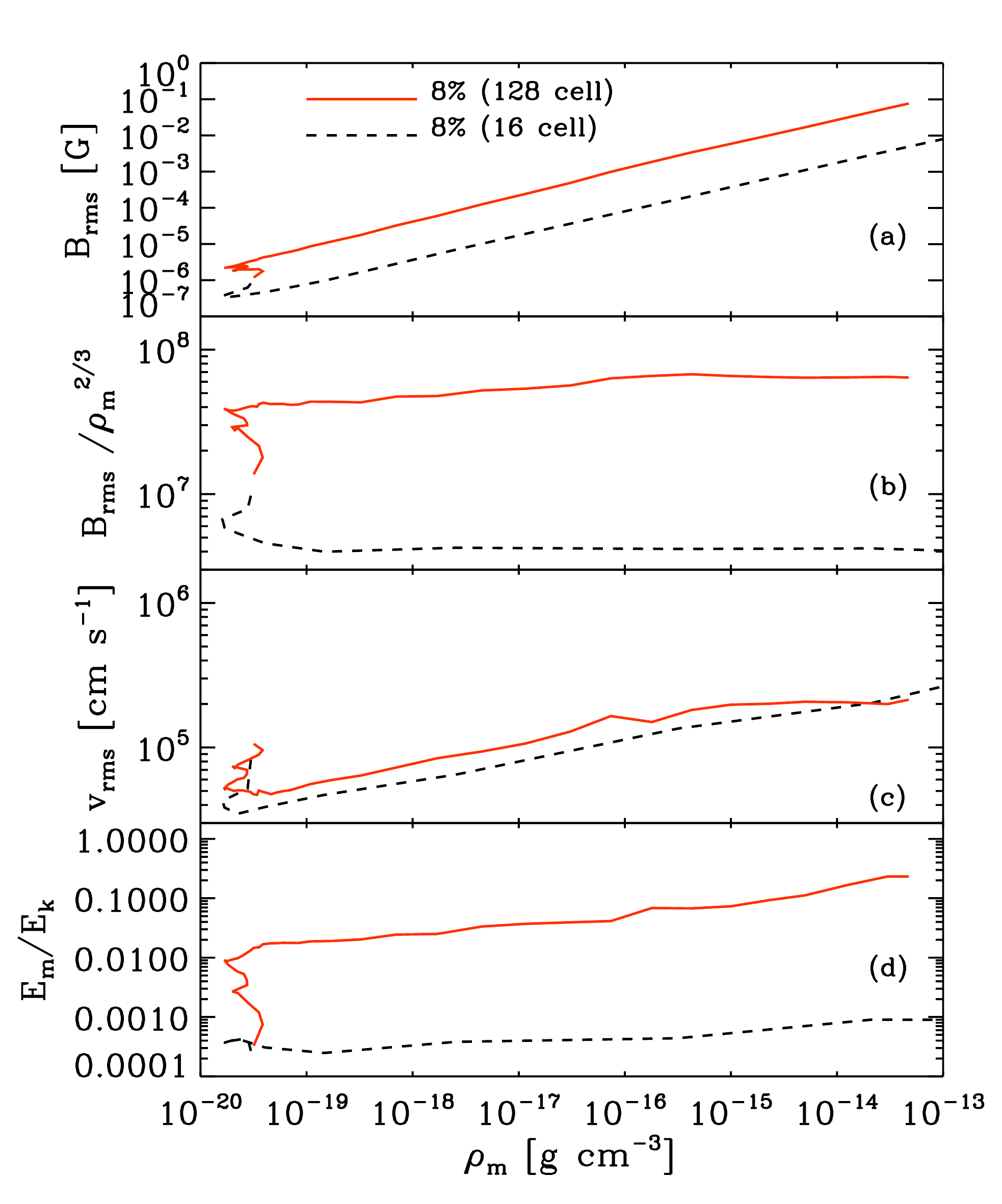}}
\caption{Comparison between two runs with initial rotation of $E_{\rm rot}/|E_{\rm grav}| = 8\%$, 
where the Jeans length was resolved by 128 cells (solid line) and 16 cells (dashed line), respectively.  
In comparison to our highest resolution simulation (128 cells), this leads to an order of magnitude difference 
in the magnetic field amplification. No dynamo amplification occurs in the 16 cell run as is evident from 
panel b (see also Papers I and II for a more detailed study of the resolution effects). 
\label{allcomb_rot_mdens_2}}
\end{figure*}

To make a direct comparison of how the collapse and field amplification are affected by rotation, we compare 
two simulations with the same value of the $\beta$ parameter, but at two different resolutions, one at $16$ cells 
and the other at a resolution of $128$ cells. The utility of this approach lies in the fact that when the local Jeans 
length is resolved by only $16$ cells, the small-scale dynamo is not excited (as turbulent motions are under-resolved) 
and the magnetic field is only amplified by gravitational compression in a rotating gas cloud. The comparison is
shown in Fig.~\ref{allcomb_rot_mdens_2}, between a 16 cell and a 128 cell run both having $\beta=8\%$ 
initial uniform rotation. For the 16 cells run, the field amplification arises purely due gravitational compression 
of the field lines. This leads to an order of magnitude difference in the evolution of the rms magnetic field (panel (a) 
in Fig.~\ref{allcomb_rot_mdens_2} ). However, as explained earlier, the growth rate of the small-scale dynamo 
depends on the resolution. Thus, an order of magnitude difference in this case should be taken as a lower limit.
The plot in panel (b) confirms the absence of any small-scale dynamo for the 16 cell run as the 
$B_{\rm rms}/\rho_{\rm m}^{2/3}$ curve at first decays and then stays roughly constant while for the 128 cell run, 
there is an initial increase in $B_{\rm rms}/\rho_{\rm m}^{2/3}$ followed by saturation at a value 
$B_{\rm rms}/\rho_{\rm m}^{2/3}\sim 6\times10^{7}$. Also, from panel (d), we find that while the ratio of 
$E_{\rm m}/E_{\rm k}$ increases rapidly to values $\sim 10^{-2}$ for the 128 cell run, there is no corresponding 
increase for the 16 cell run. We refer the reader to Papers I and II for a more detailed analysis of resolution 
effects. 

\section{Summary and Conclusions}

In this paper, we have presented a detailed study of the influence of initial conditions on 
the gravitational collapse and magnetic field amplification of a dense gas cloud. We chose 
initial and environmental conditions that are reminiscent of the conditions in primordial mini-halos
that lead to the formation of the first stars in the Universe. We purposely chose the initial strength 
of the magnetic field to be $\sim 1\, \mkG$ with an $E_{\rm m}/E_{\rm k} \sim 10^{-4}$
to capture the saturation of the small-scale dynamo. In this respect, this study goes beyond our 
earlier studies (Papers I and II) where we only considered the kinematic regime of the dynamo 
growth. Since dynamo amplification of the magnetic field occurs due to turbulent fluid motions, we 
varied the strength and injection scale of the initial turbulence in this study. In addition, we have
also studied the influence of initial uniform rotation on the collapse and magnetic field amplification. 
The general behavior of the magnetic field amplification presented here reveals two distinct phases: first, 
the total magnetic field is amplified by both the gravitational compression and the small-scale 
dynamo. Later, the small-scale dynamo saturates after which the field amplification is driven only 
by compression. Our main results obtained from the systematic study are summarised as follows:
\begin{itemize}

\item We started by exploring the effect of varying the initial strength of the turbulent velocity. The parameter
range we explored concerns an initial $v_{\rm rms}/c_{\rm s} = 0.2, 0.4, 1.0, 2.0$ and $4.0$. The total 
magnetic field in these simulations is amplified by more than 6 orders of magnitude, reaching peak values 
of $\sim 1\,{\rm G}$ at densities of $\sim 10^{-12}\,\rm{g}\,\,{\rm cm}^{-3}$. Initially, the small-scale dynamo 
provides additional field amplification to the regular amplification occurring due to gravitational compression. 
Later, we find that the small-scale dynamo saturates. In almost all the cases, except when the initial 
$v_{\rm rms} = 4\,c_{\rm s}$, the amplification of the magnetic field by the small-scale dynamo continues till 
$\sim \tau = 10$. This corresponds to the kinematic phase of the dynamo. In this phase, 
$B_{\rm rms}/\rho_{\rm m}^{2/3}$ attains a peak value in the range $\sim (2 - 6) \times 10^{7}$ (with respect 
to the initial value) in the density range $\rho_{\rm m}\sim(10^{-13} - 10^{-10})\,\rm{g}\,\,{\rm cm}^{-3}$. The 
growth rate of $B_{\rm rms}/\rho_{\rm m}^{2/3}$ at first increases with the increase in initial subsonic Mach 
numbers and then decreases in the supersonic regime. \citet{Federrath+11b}, also report a similar behavior 
with Mach number in forced turbulence-in-a-box simulations of small-scale dynamo action. Beyond 
$\rho_{\rm m}\sim 10^{-14}\,\rm{g}\,\,{\rm cm}^{-3}$, $B_{\rm rms}/\rho_{\rm m}^{2/3}$ shows a distinct 
change in slope indicating that the dynamo reaches saturation. The total field amplification in the saturated 
phase is now only due to gravitational compression. The magnetic energy still shows a 
mild increase reaching values of $0.2 - 0.4$ of the equipartition strength for the parameter range we explore. 
Thus, $B_{\rm rms}/\rho_{\rm m}^{2/3}$ is a more suitable indicator of dynamo saturation in self-gravitating 
systems. \\

\item Next, we explored the effect of varying the initial injection scale of turbulence. We compared two 
simulations with $l_{\rm inj} = 0.17\,\lambda_{\rm J}$ and $0.7\,\lambda_{\rm J}$. With a smaller injection 
scale, the turbulence decays faster as more kinetic energy has to be initialized on the smaller scales to get 
the same overall turbulent Mach number. The smaller scales are therefore dissipated more quickly resulting 
in the early collapse by $\tau\sim 1$ compared to the other run where runaway collapse sets in at $\tau\sim 4$. 
The kinematic phase of the dynamo extends to $\rho_{\rm m}\sim 10^{-12}\,\rm{g}\,\,{\rm cm}^{-3}$ for 
$l_{\rm inj} = 0.17\,\lambda_{\rm J}$ and up to $\sim 10^{-15}\,\rm{g}\,\,{\rm cm}^{-3}$ 
for $l_{\rm inj} = 0.7\,\lambda_{\rm J}$. Saturation of the dynamo is once again clearly illustrated by the change 
in slope of $B_{\rm rms}/\rho_{\rm m}^{2/3}$ compared to the evolution of $E_{\rm m}/E_{\rm k}$. 
For $l_{\rm inj} = 0.17\,\lambda_{\rm J}$, this occurs from a density $\sim 10^{-10}\,\rm{g}\,\,{\rm cm}^{-3}$
while for $l_{\rm inj} = 0.7\,\lambda_{\rm J}$, saturation starts from $\rho_{\rm m}\sim 10^{-14}\,\rm{g}\,\,{\rm cm}^{-3}$. 
By the end of the simulation, the magnetic energy attains values in the range $0.1 - 0.3$ of the equipartition values 
for these two runs. \\

\item Finally, we explored the effect of uniform initial rotation. We considered three simulations where the 
rotation parameter $\beta$ has values of $0\%, 4\%$ and $8\%$. The dynamical evolution of the system 
shows two distinct phases: an initial turbulent decay phase followed by a runaway collapse phase where 
the turbulence gets regenerated by the gravitational collapse. The dynamo generated magnetic field 
$B_{\rm rms}/\rho_{\rm m}^{2/3}$ saturates for all the three runs when the central core density 
$\rho_{\rm m} \sim 10^{-15}\,\rm{g}\,\,{\rm cm}^{-3}$. The saturation behavior is clearly revealed in the time 
evolution of magnetic field spectra (Fig.~\ref{mag_spec}) where the peak of the spectrum gradually shifts to 
larger scales (i.e., smaller $k/k_{\rm J}$ values) at late times. Consistent with earlier theoretical predictions, 
magnetic energy attains equipartition with the kinetic energy first on smaller scales and then gradually on 
larger scales. This is shown in Fig.~\ref{emekin_spec}. 
Our resolution study further shows that the dynamo amplification leads to an order of magnitude difference in the 
magnetic field amplification. Since the dynamo amplification is resolution dependent 
(higher resolution leads to stronger field amplification), this difference should be taken as a lower bound. 
\end{itemize}

In summary, using idealized simulations, our systematic study has probed for the first time, the effect 
of different initial conditions on the dynamo amplification of the magnetic field and its saturation for the specific
case of primordial cloud collapse. Our choice of the initial field strength allowed us to probe both the kinematic 
and the saturation regime of the small-scale dynamo. Detailed analysis of the magnetic field spectra shows that 
saturation is achieved initially on smaller scales and then later on, the magnetic field saturates on larger scales. 
In all our simulations presented here, the magnetic energy continues to grow at the expense of the kinetic energy 
eventually attaining values in the range, $E_{\rm m}/E_{\rm k} \sim 0.1 - 0.4$. This range of values are roughly in 
agreement with previous analytical \citep{S99} and numerical work on small-scale dynamo action in forced 
turbulence \citep{HBD04, BS05, Federrath+11b}. For the case of a primordial cloud collapse which we explore 
here, the small-scale dynamo comes across as an efficient process to generate strong seed fields on scales 
comparable to the scale of turbulence (in our case, the local Jeans length). 
These strong seed fields could potentially lead to the generation of large-scale coherent magnetic fields on scales 
much larger than the scale of turbulence (in our case on scales $k/k_{\rm J}\ll 1$) via large-scale dynamo mechanisms 
\citep{RSS88,HWK09, JS11}. The implications of such coherent magnetic fields has been previously explored in the
works of \citet{PS89} and \citet{TB04}. Starting from seed magnetic fields such as the ones generated by a small-scale 
dynamo, \citet{TB04} show that coherent magnetic fields can be produced {\it in situ} in primordial star forming disks. 

Numerical simulations with an initially coherent magnetic field have previously shown the formation of jets and 
bipolar outflows \citep[e.g.,][]{BP06,Machida+06,BP07, MIM08}. Recent work by \citet{GCS10} find evidence of 
small-scale dynamo action in the solar surface. \citet{SL06} infer that magnetic fields can be amplified 
by the magneto-rotational instability (MRI) in the disk leading to coherent magnetic fields. The presence of a small-scale 
dynamo can potentially further amplify the seed fields in the disk. It remains to be seen what effect, if such large-scale 
magnetic fields generated during the primordial collapse, have for example - on the formation process of the first stars. 
Recent hydrodynamical calculations of first star formation \citep{Greif+11, Clark+11} have shown that the disks that 
formed around the first young stars were unstable to gravitational fragmentation, possibly leading to the formation of small 
binary and higher-order systems. Whether or not, magnetic fields generated and amplified by dynamo processes can 
influence this scenario is an open question. Magnetic fields may also have a bearing on the rotation speed of the first stars 
\citep{SBL11}. Moreover, magnetic fields generated by dynamo processes in the primordial universe and ejected in outflows 
will have implications for the formation of the second generation of stars and the first galaxies. More detailed investigations 
including other relevant physics and additional processes like primordial chemistry and cooling, non-ideal MHD
and radiative feedback effects are needed in the near future to address these possibilities.

\section*{Acknowledgements}
S.S. thanks the German Science Foundation (DFG) for financial support via the priority 
program 1177 ÒWitnesses of Cosmic History: Formation and Evolution of Black Holes, 
Galaxies and their EnvironmentÓ (grant KL 1358/10). S.S further thanks IUCAA and RRI for 
support and hospitality where this work was completed. C. F.~received funding from the Australian
Research Council (grant DP110102191) and from the European Research Council (FP7/2007-2013
Grant Agreement no.~247060). D.~R.~G.~S thanks for funding through the SPP 1573 (project number 
SCHL~1964/1-1) and the SFB 963/1 {\em Astrophysical Flow Instabilities and Turbulence}. C.F.,~R.B., 
and R.S.K.~acknowledge subsidies from the Baden-W{\"u}rttemberg-Stiftung (grant P-LS-SPII/18) and 
from the German Bundesministerium f{\"u}r Bildung und Forschung via the ASTRONET project STAR 
FORMAT (grant 05A09VHA). R.B.~acknowledges funding by the Emmy-Noether grant (DFG) BA~3706. 
Supercomputing time at the Forschungszentrum J\"ulich via grant hhd142 and hhd20 and at the Leibniz 
Rechenzentrum via grant no. pr42ho and pr32lo is gratefully acknowledged. The software used in this work 
was in part developed by the DOE-supported ASC/Alliance Center for Astrophysical Thermonuclear Flashes 
at the University of Chicago.
%\bibliographystyle{apj}
%\bibliography{astro}

%%%%%%%%%%%%%%%%%%%%%%%%%%%%%%%%%%%%%%%%%%%%%%%%%%
%%%%%%%%%%%%%%%%%%%%%%%%%%%%%%%%%%%%%%%%%%%%%%%%%%%
%%%%%%%%%%%%%%%%%%%%%%%%%%%%%%%%%%%%%%%%%%%%%%%%%%%
%%%%%%%%%%%%%%%%%%%%%%%%%%%%%%%%%%%%%%%%%%%%%%%%%%%%%%
%%%%%%%%%%%%%%%%%%%%%%%%%%%%%%%%%%%%%%%%%%%%%%%%%%%%%%
\befone 
\showfour{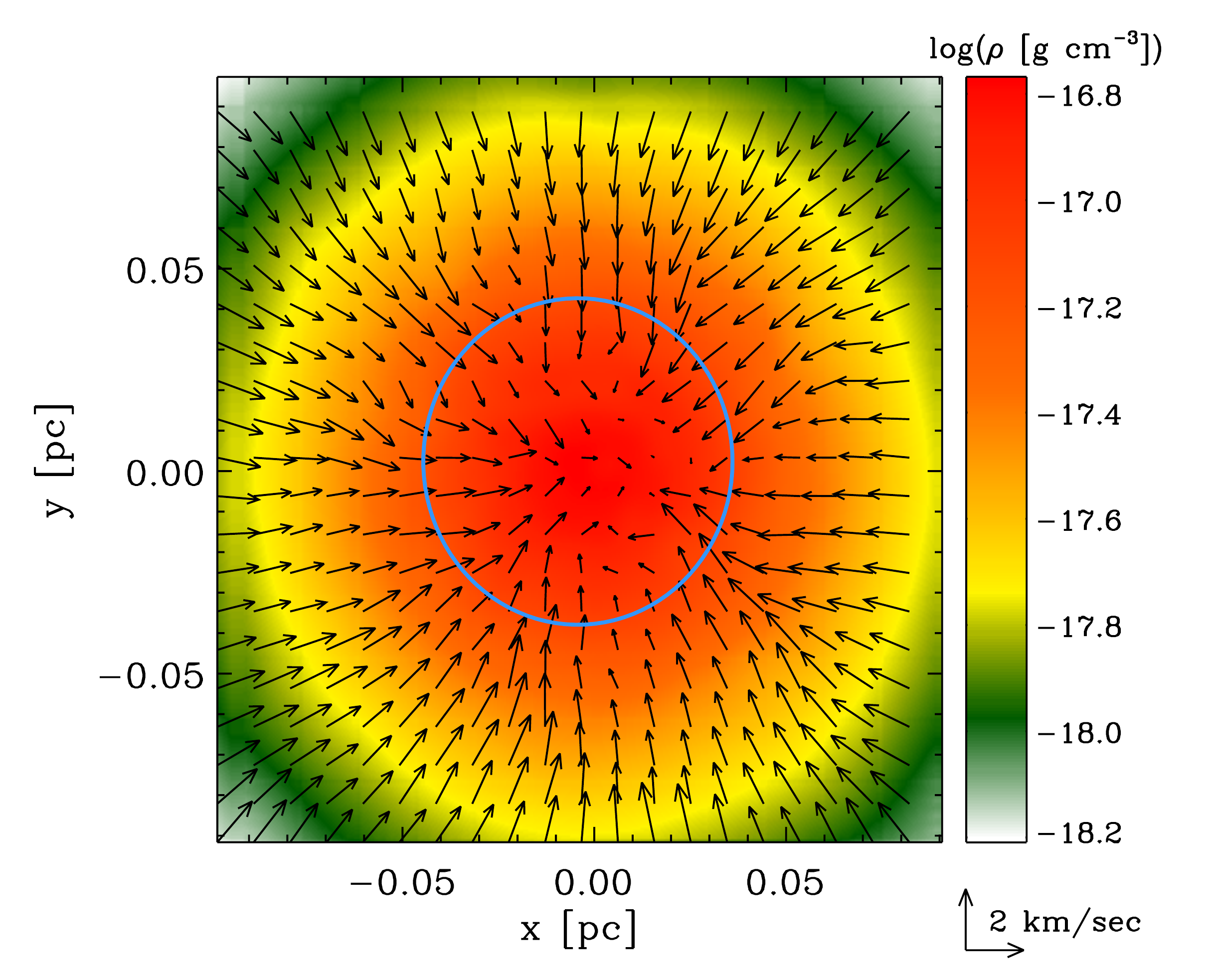}
          {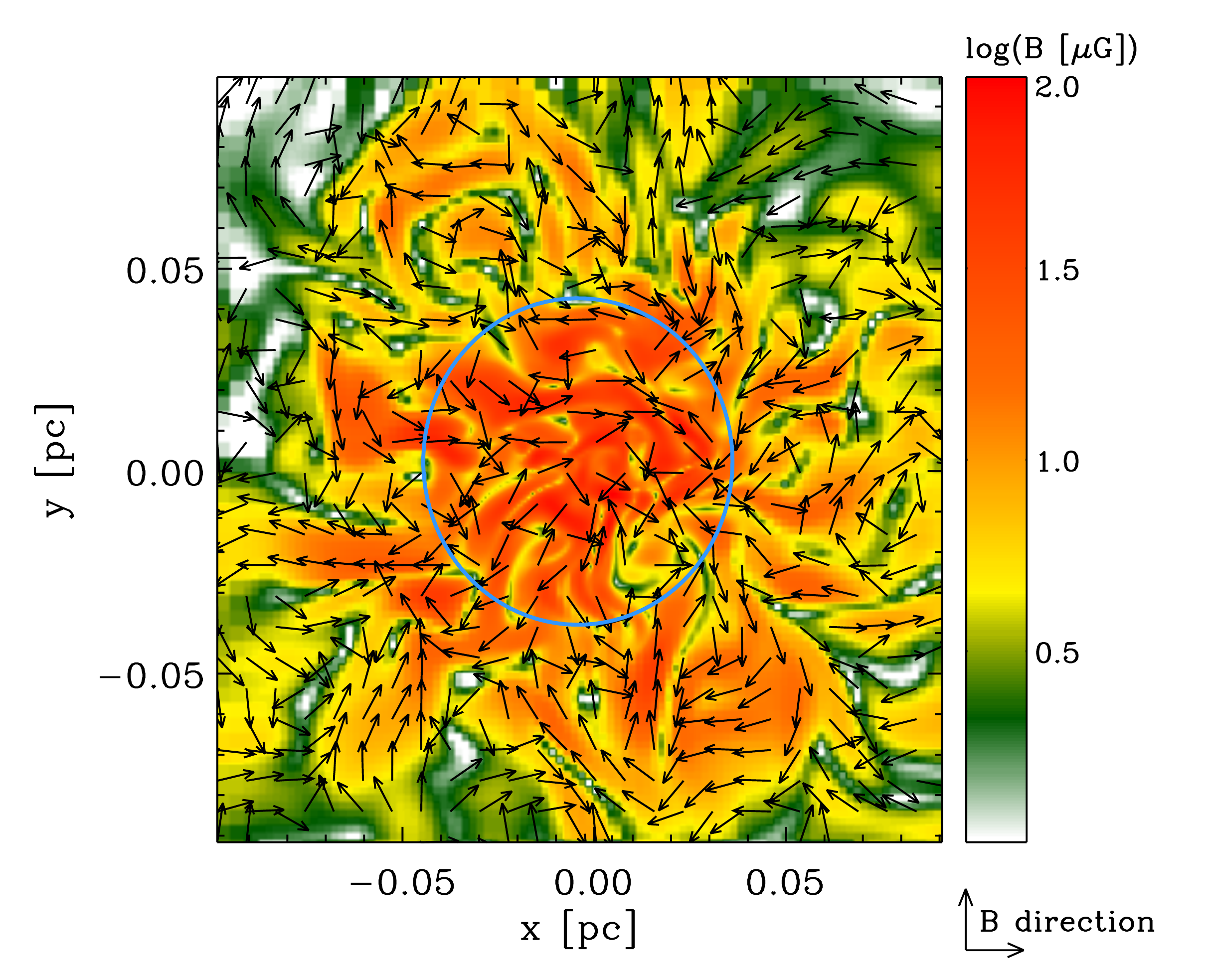}
          {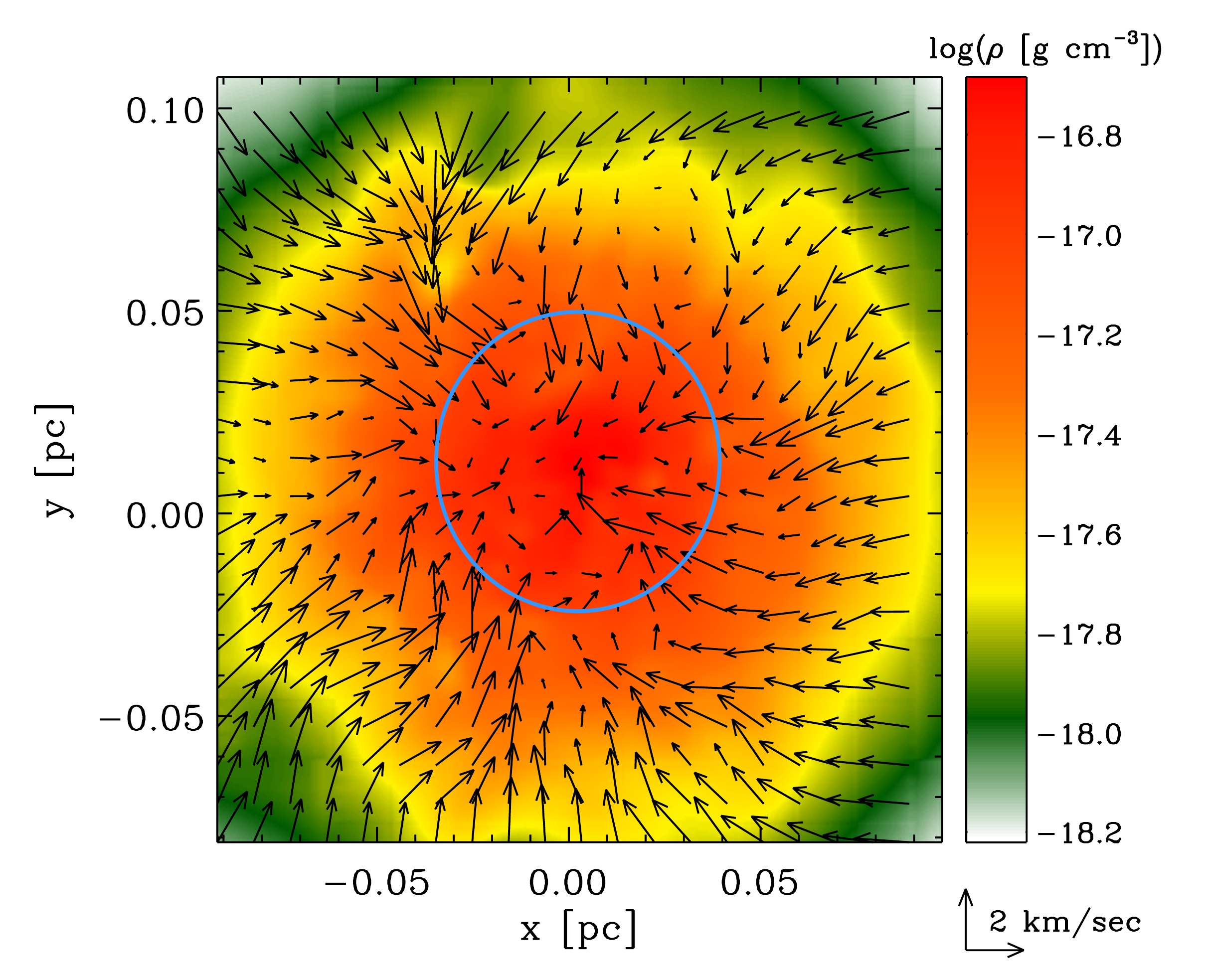}
          {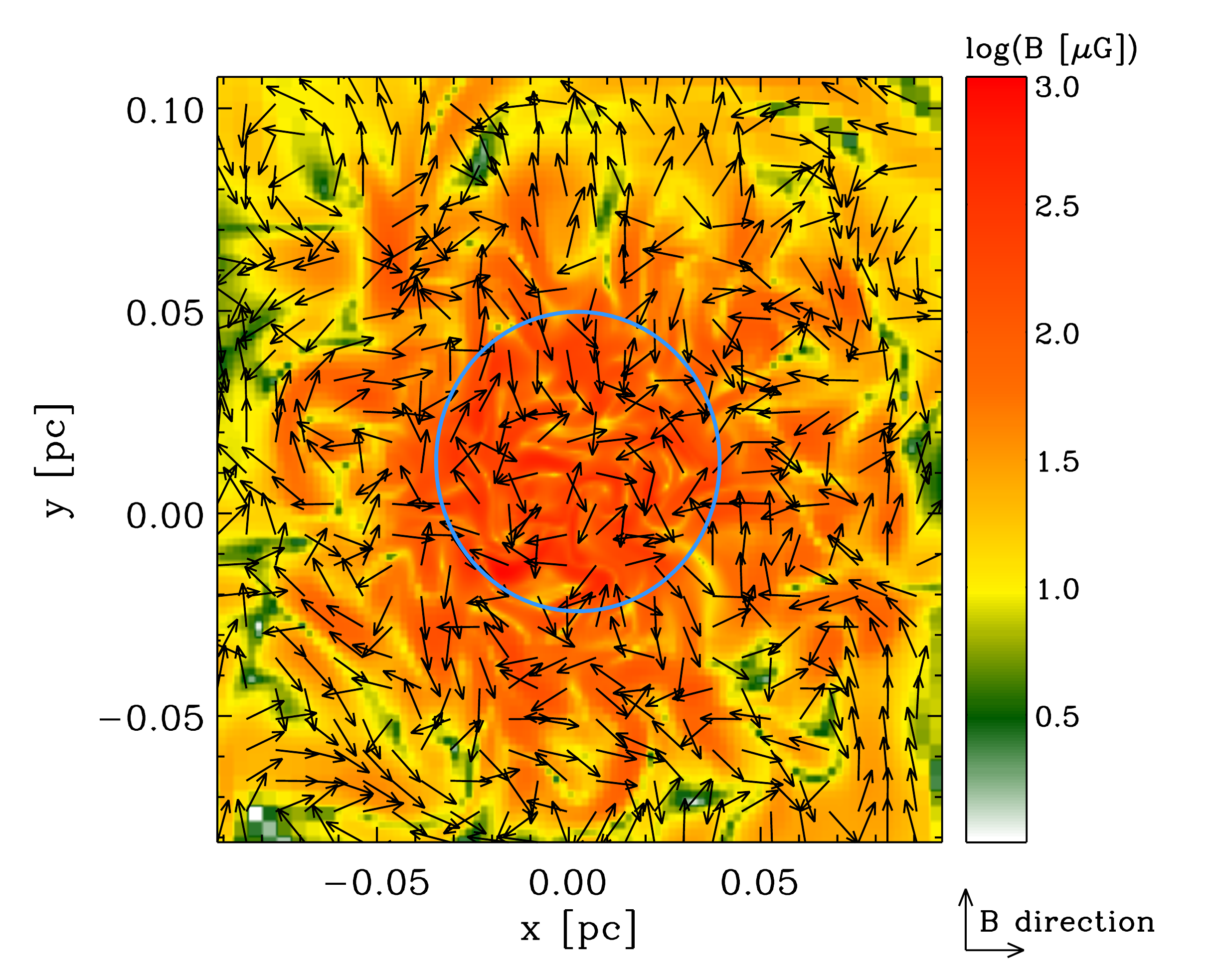}
\caption{Two-dimensional slices of the density and the total magnetic 
field for runs with different initial injection scale at a time when the central
core density is $\sim 10^{-17}\,\rm{g}\,\,{\rm cm}^{-3}$. The upper row is for 
$\l_{\rm inj}/\lambda_{\rm J}=0.17$ while the lower row corresponds to
$\l_{\rm inj}/\lambda_{\rm J}=0.7$. In the density snapshots, the arrows 
denote the velocity vectors, while those in the total magnetic field snapshots
denote the direction of the local magnetic field. The run with $\l_{\rm inj}/\lambda_{\rm J}=0.17$
shows less turbulent motions inside the Jeans volume compared to the run with 
$\l_{\rm inj}/\lambda_{\rm J}=0.7$. The magnetic field attains a peak value of 
$\sim 10^{-4}\,{\rm G}$ for $\l_{\rm inj}/\lambda_{\rm J}=0.17$ and $\sim 10^{-3}\,{\rm G}$ 
for $\l_{\rm inj}/\lambda_{\rm J}=0.7$ for the same central density of 
$\rho_{\rm m} \sim 10^{-17}\,\rm{g}\,\,{\rm cm}^{-3}$.}
\label{lj_slices}
\eefone
%%%%%%%%%%%%%%%%%%%%%%%%%%%%%%%%%%%%%%%%%%%%%%%%%%%%
%%%%%%%%%%%%%%%%%%%%%%%%%%%%%%%%%%%%%%%%%%%%%%%%%%%
%%%%%%%%%%%%%%%%%%%%%%%%%%%%%%%%%%%%%%%%%%%%%%%%%%%
%%%%%%%%%%%%%%%%%%%%%%%%%%%%%%%%%%%%%%%%%%%%%%%%%%%%
%%%%%%%%%%%%%%%%%%%%%%%%%%%%%%%%%%%%%%%%%%%%%%%%%%%%%%
%%%%%%%%%%%%%%%%%%%%%%%%%%%%%%%%%%%%%%%%%%%%%%%%%%%%
%%%%%%%%%%%%%%%%%%%%%%%%%%%%%%%%%%%%%%%%%%%%%%%%%%
\befone 
\showfour{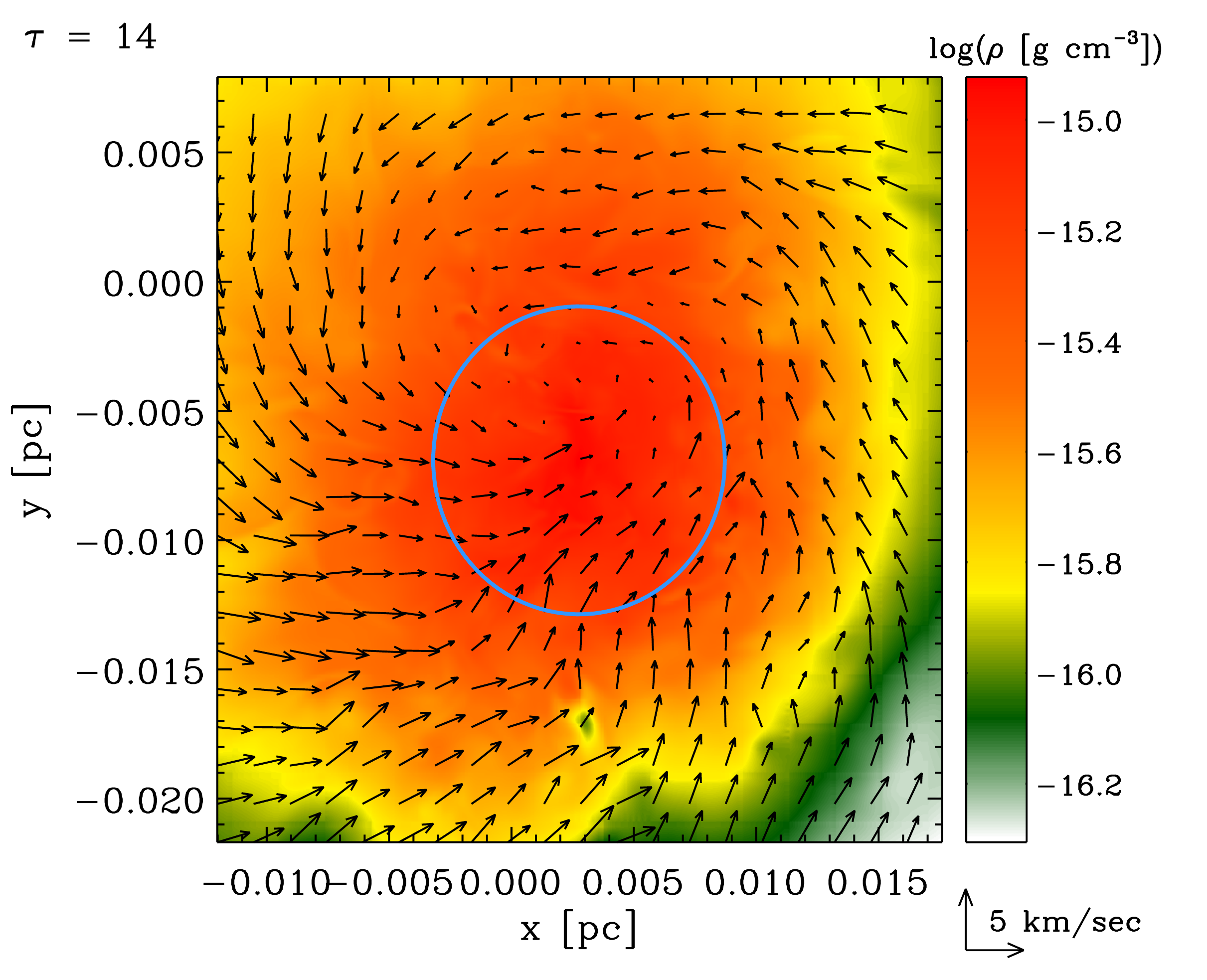}
          {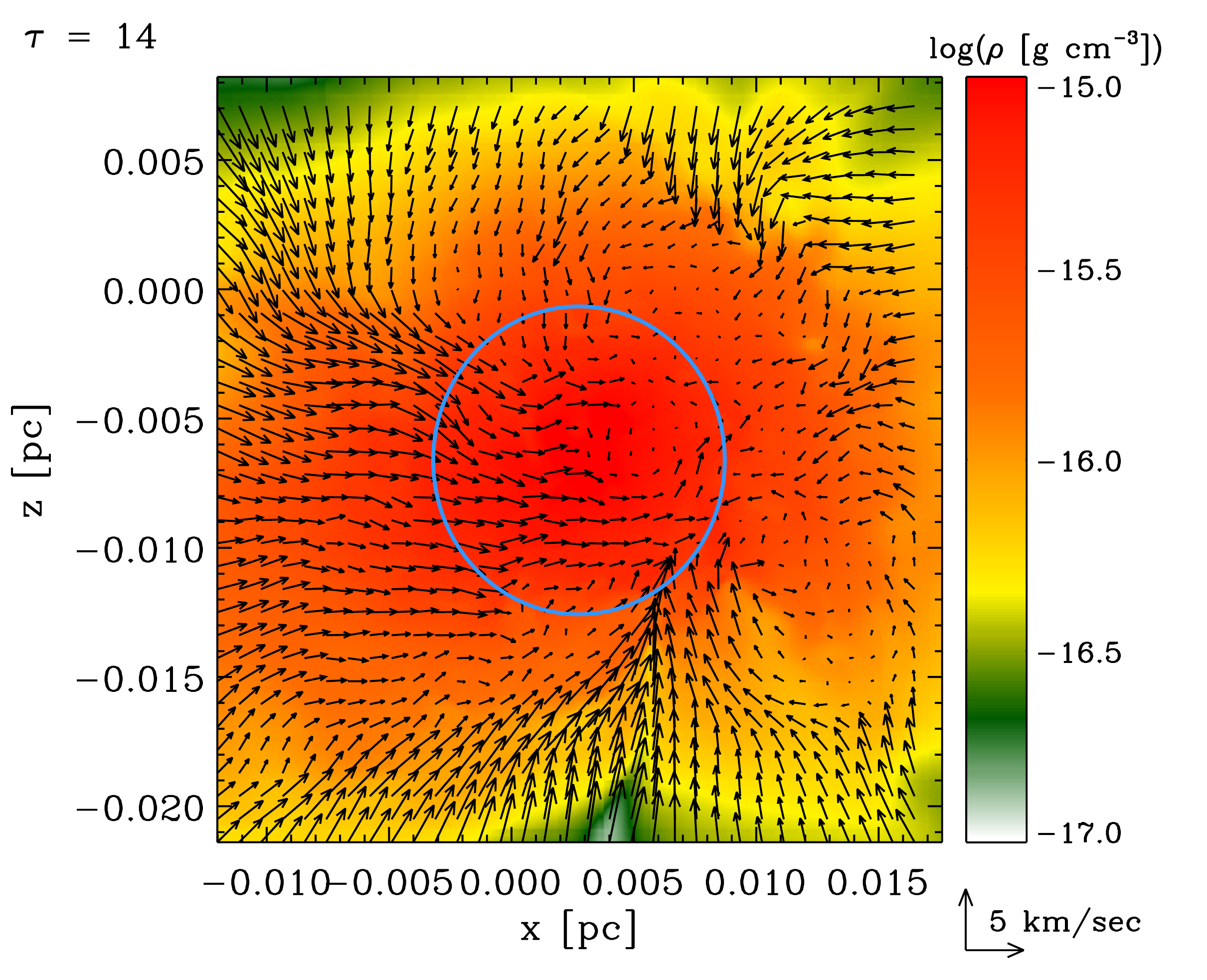}
          {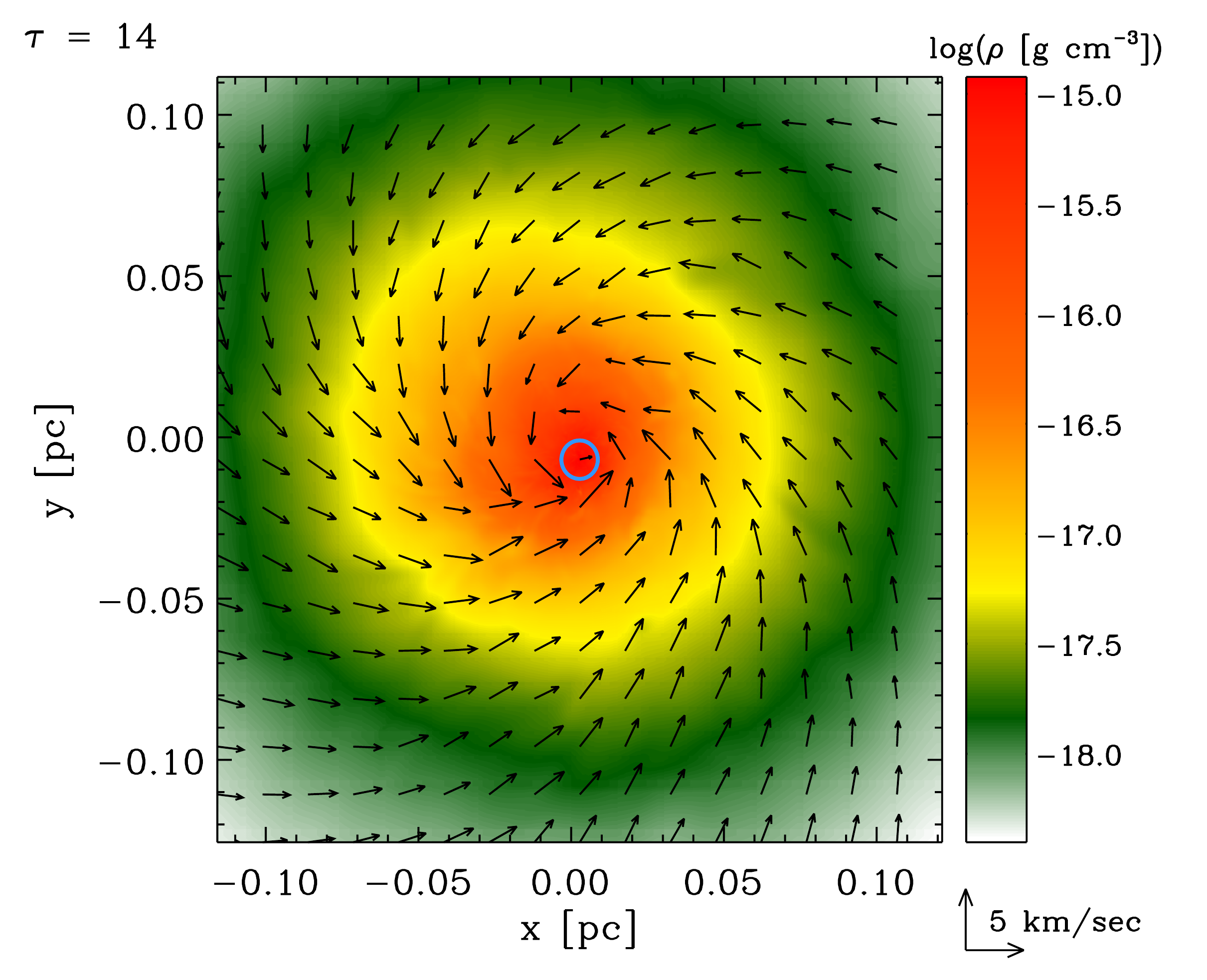}
          {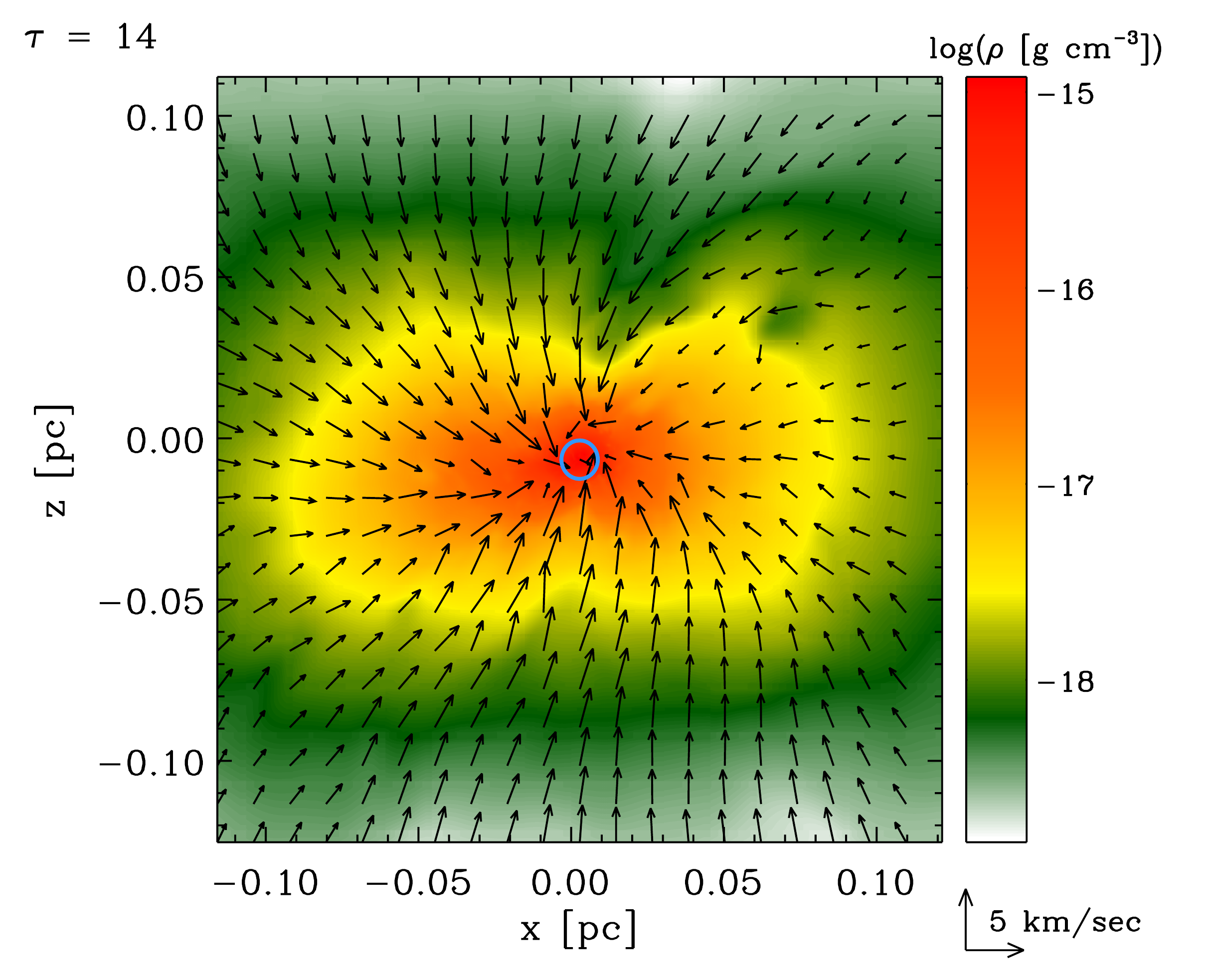}
\caption{Morphology of the cloud: Two-dimensional slices of the density for a run with $8\%$ 
initial uniform rotation through the center of the collapsing core at $\tau=14$. The top row 
corresponds to zoomed-in slices in the $x$-$y$ plane and $x$-$z$ plane, while the bottom 
row corresponds to zoomed-out slices in the $x$-$y$ plane and in the $x$-$z$ plane. The 
arrows denote the velocity field vectors, while the circle denotes the local Jeans volume. 
The zoomed out $x$-$z$ slice on the bottom row shows the gradual flattening of the collapsed 
BE sphere due to the initial rotation.}
\label{dens_slices}
\eefone
%%%%%%%%%%%%%%%%%%%%%%%%%%%%%%%%%%%%%%%%%%%%%%%%%%%%%%
\befone 
\showfour{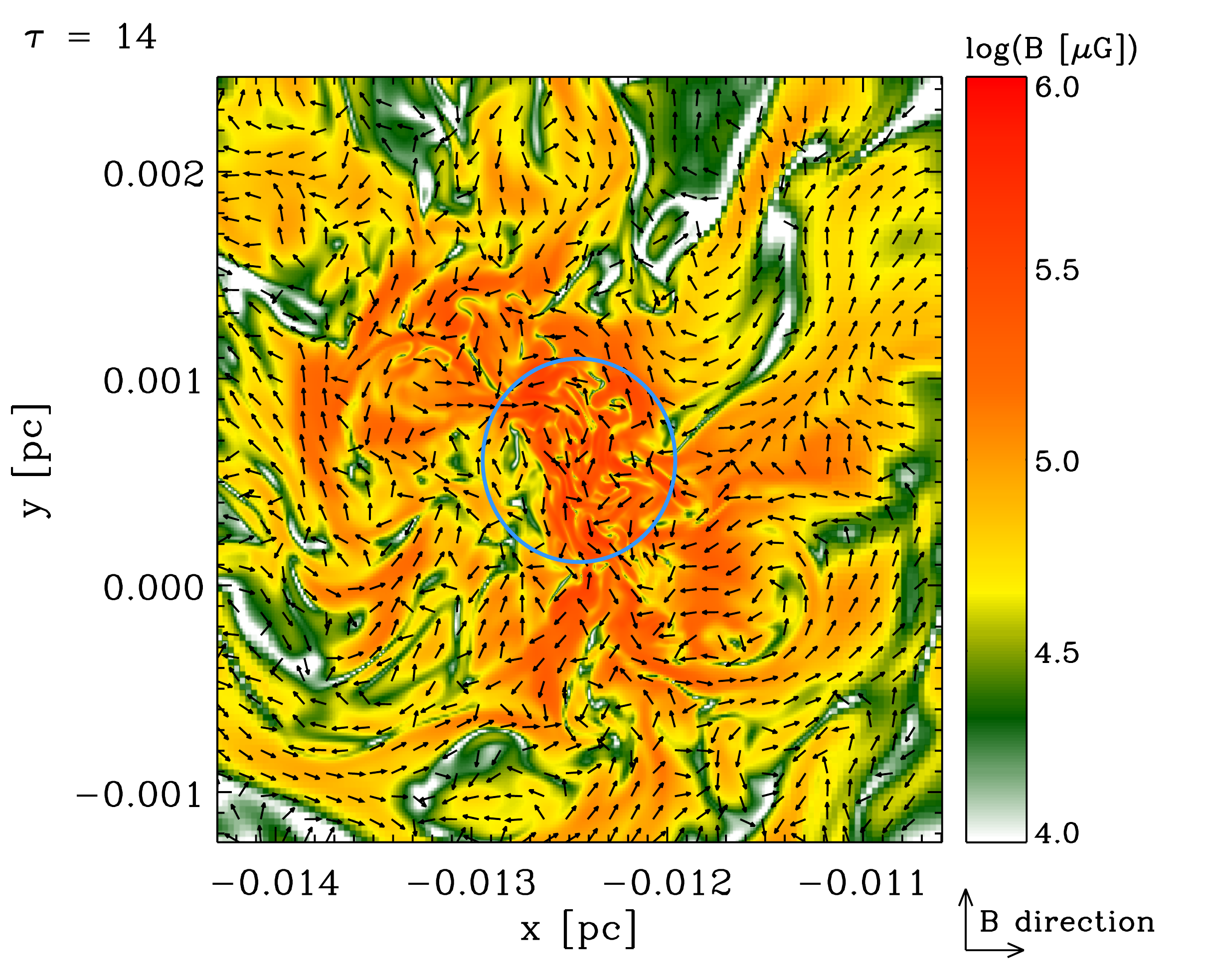}
          {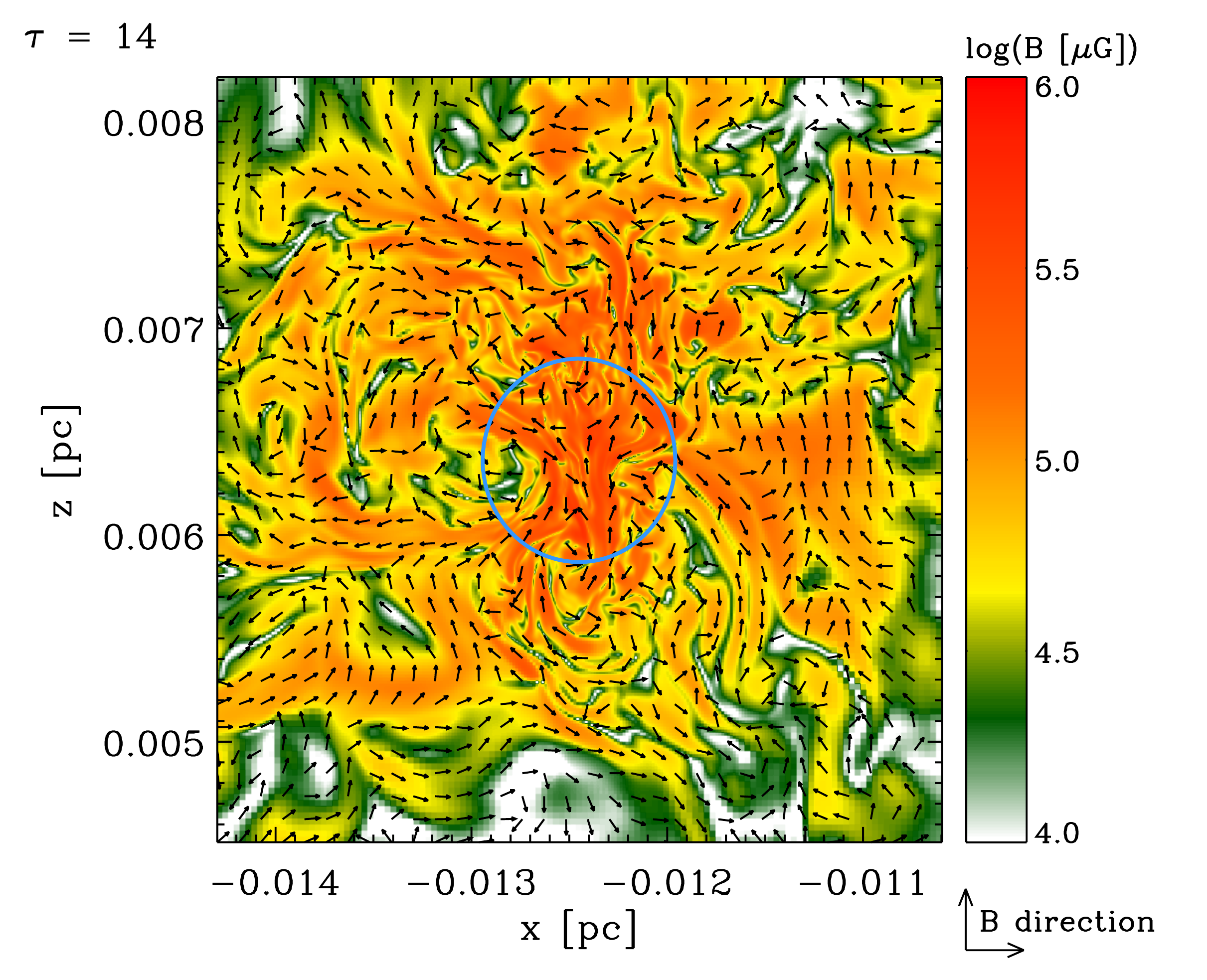}
          {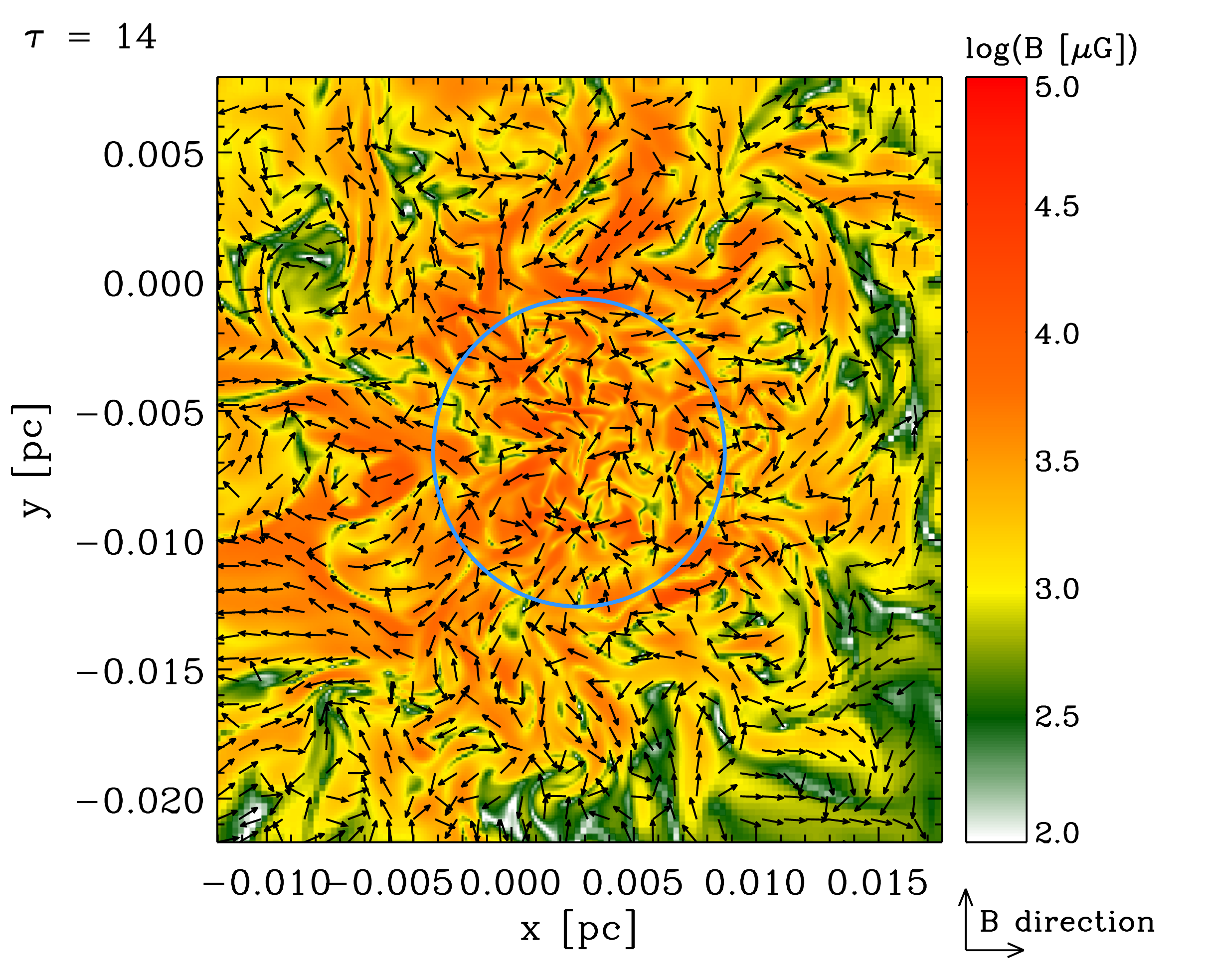}
          {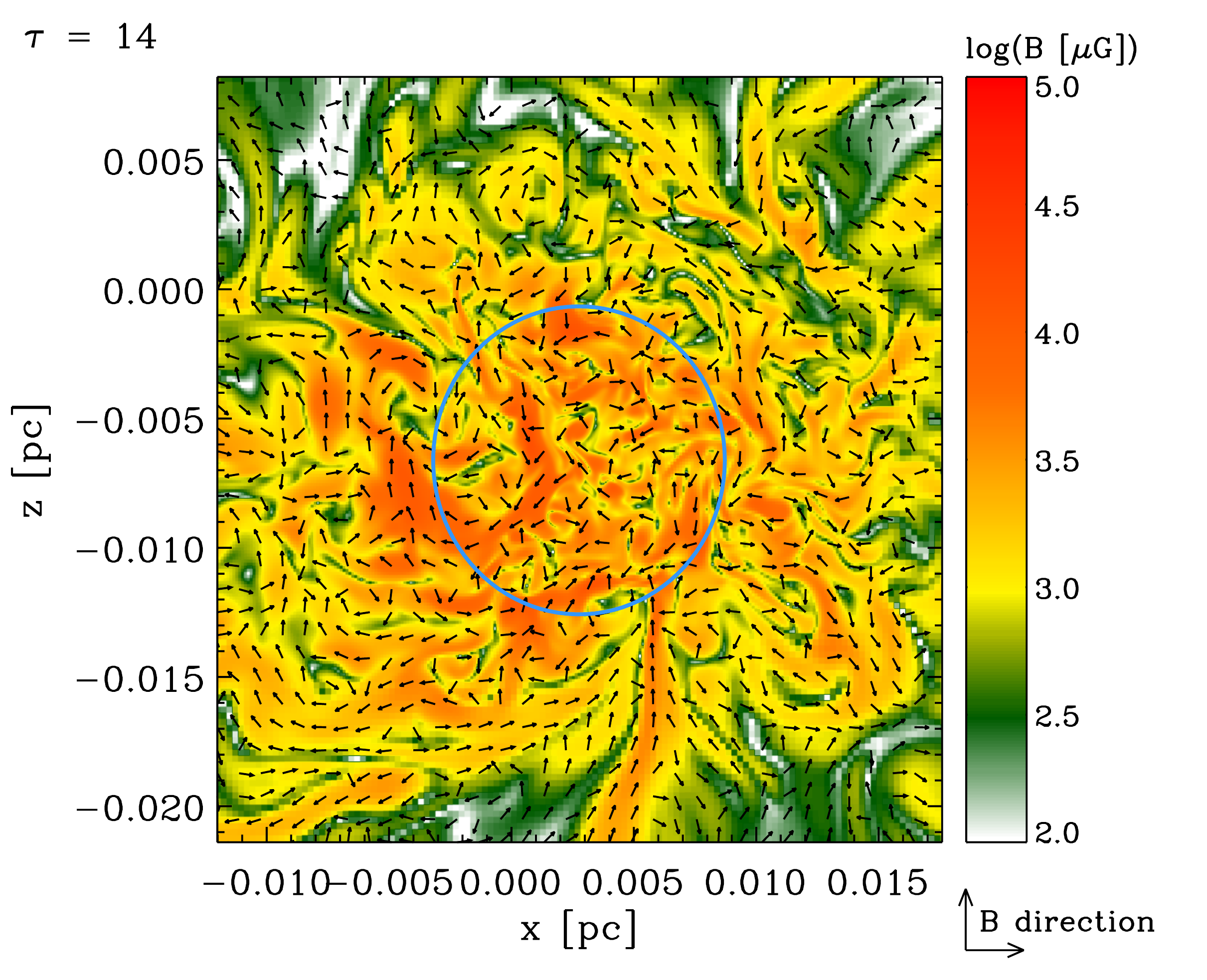}
\caption{Magnetic field structures: Two-dimensional slices of the total magnetic field through the 
center of the collapsing sore at  $\tau=14$. The upper row shows the $x$-$y$ and the $x$-$z$ 
plane snapshots for a run with no initial uniform rotation. The lower row shows the $x$-$y$ 
and the $x$-$z$ plane snapshots for a run with $8\%$ uniform rotation. As before, the 
circles depict the local Jeans volume. The arrows denote the direction of the local 
magnetic field.}
\label{totmag_slices}
\eefone
%%%%%%%%%%%%%%%%%%%%%%%%%%%%%%%%%%%%%%%%%%%%%%%%%%
%%%%%%%%%%%%%%%%%%%%%%%%%%%%%%%%%%%%%%%%%%%%%%%%%%
%%%%%%%%%%%%%%%%%%%%%%%%%%%%%%%%%%%%%%%%%%%%%%%%%%%%

\label{lastpage}
\end{document}